\documentclass[fleqn,usenatbib]{mnras}

\usepackage{newtxtext,newtxmath}

\usepackage[T1]{fontenc}

\DeclareRobustCommand{\VAN}[3]{#2}
\let\VANthebibliography\thebibliography
\def\thebibliography{\DeclareRobustCommand{\VAN}[3]{##3}\VANthebibliography}

\usepackage{graphicx}	
\usepackage{amsmath}	
\usepackage{chemformula}
\usepackage{chemmacros}
\usepackage{xspace}

\newcommand{\fc}{\textsc{FastChem}\xspace}
\newcommand{\fcc}{\textsc{FastChem Cond}\xspace}

\title[\texttt{FastChem\;Cond}]{\texttt{FastChem\;Cond}: Equilibrium chemistry with condensation and rainout for cool planetary and stellar environments}

\author[D. Kitzmann, J. W. Stock and A. B. C. Patzer]{
Daniel Kitzmann,$^{1}$\thanks{E-mail: daniel.kitzmann@unibe.ch}
Joachim W. Stock,
A. Beate C. Patzer$^{2}$
\\
$^{1}$Center for Space and Habitability, University of Bern, Gesellschaftsstrasse. 6, 3012 Bern, Switzerland\\
$^{2}$Zentrum f\"ur Astronomie und Astrophysik (ZAA), Technische Universit\"at Berlin (TUB), Hardenbergstr. 36, 10623 Berlin, Germany
}

\date{Accepted 2023 November 12. Received 2023 November 09; in original form 2023 September 04}
\pubyear{2023}

\begin{document}
\label{firstpage}
\pagerange{\pageref{firstpage}--\pageref{lastpage}}
\maketitle

\begin{abstract}
Cool astrophysical objects, such as (exo)planets, brown dwarfs, or asymptotic giant branch stars, can be strongly affected by condensation. Condensation does not only directly affect the chemical composition of the gas phase by removing elements but the condensed material also influences other chemical and physical processes in these objects. This includes, for example, the formation of clouds in planetary atmospheres and brown dwarfs or the dust-driven winds of evolved stars.
In this study we introduce \fcc, a new version of the \fc equilibrium chemistry code that adds a treatment of equilibrium condensation. Determining the equilibrium composition under the impact of condensation is complicated by the fact that the number of condensates that can exist in equilibrium with the gas phase is limited by a phase rule. However, this phase rule does not directly provide information on which condensates are stable. As a major advantage of \fcc is able to automatically select the set stable condensates satisfying the phase rule. Besides
the normal equilibrium condensation, \fcc can also be used with the rainout approximation that is commonly employed in atmospheres of brown dwarfs or (exo)planets. \fcc is available as open-source code, released under the GPLv3 licence. In addition to the C++ code, \fcc also offers a Python interface. Together with the code update we also add about 290 liquid and solid condensate species to \fc.

\end{abstract}


\section{Introduction}

Condensation is an important process in astrophysical objects that strongly affect the chemical composition of their gas phases. Typically, it can become very important at temperatures lower than roughly 2000 K, when the first, highly refractory species usually start to condensate out of the gas phase. Not only are the abundances of molecules directly affected by this process \citep[e.g.][]{Visscher2005ApJ...623.1221V}, but the presence of condensing species also play an important role in other physical and chemical processes in such objects. 

For example, condensation in atmospheres of (exo)planets and brown dwarfs can lead to the formation of cloud particles that strongly impact the spectral appearance and temperature structures of these objects \citep[e.g.][]{Burrows1999ApJ...512..843B, Ackerman2001ApJ...556..872A, Helling2006A&A...455..325H, Kitzmann2011A&A...531A..62K, Morley2012ApJ...756..172M, Lee2016A&A...594A..48L, Gao021JGRE..12606655G}. Moreover, the formation of dust grains in the atmospheres of asymptotic giant branch (AGB) stars, on the other hand, results in a massive stellar wind that finally leads to the loss of almost the entire stellar envelope to the interstellar medium over time \citep[e.g.][]{Hoefner1998A&A...340..497H, Winters2000A&A...361..641W, Gail2013pccd.book.....G}. Other environments where condensation is an important process are protostellar nebulae. For example, condensation sequences in the solar nebula are studied to determine the composition of planet building blocks and meteorites \citep[e.g.][]{Fegley1985E&PSL..72..311F, Palme1990E&PSL.101..180P,Lodders1993E&PSL.117..125L}. Thus, the treatment of condensation and its impact on the gas phase composition in an equilibrium chemistry code is a fundamental necessity to achieve a deeper understanding of cooler astrophysical environments. An extensive overview of condensation in astrophysical environments and a summary of modelling efforts is discussed by, for example, \citet{Ebel2006mess.book..253E}. More detailed discussion and a historical overview of the different approaches to the condensation problem and corresponding chemistry codes is provided by \citet{Lodders1997AIPC..402..391L} or \citet{Ebel2021}.

The chemical equilibrium approximation simplifies the problem of describing condensate formation significantly, albeit at the expense of assuming local thermodynamic equilibrium. Although this assumption is not flawless, it remains reasonable for various scenarios. Nevertheless, this equilibrium is not attained in regions with very low pressures and temperatures, such as the outer circumstellar shells of AGB stars. In such an environment, for example, the timescale of the hydrodynamic gas flow is much shorter than that of chemical reactions, resulting in a state where the chemical composition essentially becomes frozen \citep{Gail2013pccd.book.....G}.

Yet, despite the inadequacy of chemical equilibrium condensation in certain environments, its computation remains a crucial necessity. Chemical equilibrium calculations establish a fundamental benchmark, serving as a foundation to extend computations to encompass kinetic effects on chemistry. Without knowledge of equilibrium conditions for reference, it becomes unfeasible to ascertain whether a system is in or out of equilibrium, or to determine the degree of deviation from equilibrium \citep{Ebel2021}.

Equilibrium chemistry calculations are typically done either via the law of mass action in combination with the element conservation equations \citep{Brinkley46, Brinkley47} or by using a Gibbs free energy minimisation approach \citep{Whi57,Whi58}. Both formalisms yield the same solution but are numerically different \citep{Zel60, Zel68, smithchemical, vanZeggeren1970}. The first forms a system of non-linear equations while the latter is a global optimisation problem. On the one hand, examples of computer codes that use the law-of-mass-action approach are \textsc{CONDOR} \citep{Lodders1993E&PSL.117..125L}, \textsc{GGChem} \citep{Woitke2018A&A...614A...1W}, or \fc \citep{Stock2018MNRAS.479..865S,Stock2022MNRAS.517.4070S}. The Gibbs free energy minimisation method, on the other hand, is implemented for example in the RAND method \citep{Whi58}, the \textsc{NASA-CEA} code \citep{Gordon1994, McBride96}, SOLGAS \& SOLGASMIX \citep{Eriksson1971, Eri75}, or \textsc{STANJAN} \citep{Rey86}. A more detailed overview and descriptions of the different numerical methods used to compute the chemical equilibrium composition can be found in, for example, \citet{smithchemical} or \citet{vanZeggeren1970}.

Determining the equilibrium composition in a case where condensation occurs is quite more elaborate than a pure gas-phase calculation since only a limited number of condensates can co-exist in equilibrium with the gas phase at the same time according to an extension of Gibbs' famous phase rule \citep{Gib76,Den55}. In particular, it is in general not a priori obvious which species are stable. The phase rule only limits the number of stable condensates but doesn't restrict the potential combinations of condensates that can be present. This is especially problematic at low temperatures when many different species could in principle condense and, thus, the number of possible combinations could be very large. 

Equilibrium condensation is therefore tpyically treated as a two-step process that includes solving for the combined gas-condensed phase chemical composition as well as finding the set of stable condensates. The latter step is often done in an iterative way by adding and/or removing potential condensates from a list of stable condensate candidates until the phase rule is satisfied and element conservation by the gas and condensed phase species is achieved \citep[e.g.][]{Gordon1994, Woitke2018A&A...614A...1W}. Such iterative methods, however, become very challenging once the number of potential condensates is large. In such a case, additional a priori information on condensation sequences might be required to limit the number of combinations. While an iterative way is an established approach to solve this problem in astrophysical contexts, alternative methods do also exist \citep[e.g.][]{Leal2016AdWR...96..405L}.

The first two version of our equilibrium chemistry code \fc focused solely on the gas phase chemistry. In the original publication \citep{Stock2018MNRAS.479..865S} the basic semi-analytical approach that our chemistry code uses was introduced but was strictly limited to chemical systems including hydrogen. In \fc version 2 \citep{Stock2022MNRAS.517.4070S} we then, among other things, adapted the formalism to treat arbitrary element compositions. 

In this study we adapt \fc to include equilibrium condensation. With an internal version number of 3.0, we refer to this updated version as \fcc. In addition to equilibrium condensation, \fcc also employs the rainout approximation as an alternative condensation description that is commonly used in (exo)planetary science or for atmospheres of brown dwarfs \citep{Marley2013cctp.book..367M}. In Section \ref{sect:method} we first briefly summarise the theoretical concepts of equilibrium condensation and the related computational challenges. Section \ref{sec:fastchem_cond} describes the numerical framework implemented in \fcc, while in Section \ref{sec:model_tests} we test our new equilibrium chemistry code on a set of different scenarios from the literature that involve condensation.

\section{Theoretical framework of equilibrium condensation}
\label{sect:method}

\subsection{Gas phase treatment}
\label{sec:gas_phase}

Below we very briefly summarise the equations describing chemical equilibrium in the gas phase using the law of mass action. For a more detailed description of the notation and formalisms used in the following we refer to \citet{Stock2018MNRAS.479..865S} and \citet{Stock2022MNRAS.517.4070S}.

The number density of gas phase species $n_i$ is given by the law of mass action:
\begin{equation}
    n_{i} = K_i(T) \prod_{j \in \mathcal E} {n_j}^{\nu_{ij}}  \ , \qquad\qquad i \in \mathcal S \setminus \mathcal E \ ,
    \label{eq:mass_action_molecules}
\end{equation}
where $n_j$ is the number density of elements $j \in \mathcal E$ in their atomic form, $K_i$ the temperature-dependent equilibrium constant of species $i$, and $\nu_{ij}$ their stoichiometric coefficients. The set $\mathcal{E}$ contains all elements, while the set $\mathcal{S}$ includes all gas phase species.

With the relative, normalised element abundances $\hat \epsilon_j$, the total number density of atomic nuclei of a given element $j \in \mathcal E$ is given by
\begin{equation}
    N_j := \hat \epsilon_j n_\mathrm{\left\langle g\right\rangle} \ ,
    \label{eq:def_nj}
\end{equation}
where the total number density of all atomic nuclei is defined by
\begin{equation}
    n_\mathrm{\left\langle g\right\rangle} := \sum_{j \in \mathcal E} n_j + \sum_{j \in \mathcal E} \sum_{i \in \mathcal S \setminus \mathcal E} \nu_{ij} n_i \ .
\end{equation}
We can then write the corresponding element conservation equation for element $j \in \mathcal E$:
\begin{equation}
    N_j = n_j +  \sum_{i \in \mathcal S \setminus \mathcal E} \nu_{ij} n_i \ .
    \label{eq:gas_element_conservation}
\end{equation}
Finally, assuming an ideal gas, the gas pressure $p_\mathrm{g}$ is given by
\begin{equation}
    p_\mathrm{g} = k_\mathrm{B} T n_g = k_\mathrm{B} T \left( \sum_{j \in \mathcal E} n_j + \sum_{i \in \mathcal S \setminus \mathcal E} n_i \right) \ ,
    \label{eq:gas_pressure}
\end{equation}
where $k_\mathrm{B}$ denotes Boltzmann's constant and $T$ the gas temperature.

Equations \eqref{eq:mass_action_molecules}, \eqref{eq:gas_element_conservation}, and \eqref{eq:gas_pressure} form a system of linear and non-linear algebraic equations for the number densities of all gas phase species. The semi-analytical computational procedure used to solve this system is described in detail by \citet{Stock2022MNRAS.517.4070S} (\textsc{FastChem 2}).

\subsection{Stoichiometric condensates}

In the following we only consider neutral, stoichiometric condensates\footnote{The term condensate covers both solids and liquids throughout this work.}. Other cases, such as solid solutions for example, are not considered here.

Let $\mathcal C = \{C_1, \ldots,  C_{\lvert \mathcal C \rvert}\}$ be the set of all considered condensates in the chemistry model, where $\lvert \mathcal C \rvert$ denotes the total number of condensates.

Unlike gas phase species described in the previous subsection, not all condensates can stably exist in equilibrium with the gas phase, but only a subset $\mathcal C_s \subset \mathcal C$. 
Depending on the chosen values for the temperature $T$, the gas pressure $p_{\mathrm g}$, and the element abundances $\hat \epsilon_j$, $\mathcal C_s$ can also be an empty set, i.e. no condensate exists. This is the case for sufficiently high temperatures, for example. Furthermore, we introduce the subset of elements $\mathcal E_s \subset \mathcal E$ that are contained in the stable condensates $\mathcal C_s$.

The stability criterion of a condensate $c$ can be expressed by using its activity $a_c$, given by the corresponding law of mass action \citep{Gail2013pccd.book.....G}:
\begin{equation}
    a_{c} = K_c(T) \prod_{j \in \mathcal E} {n_j}^{\nu_{cj}} \ , \qquad\qquad c \in \mathcal C \ ,
    \label{eq:activity_lin}
\end{equation}
where $K_c$ is the equilibrium constant for the condensate.

For numerical computations, Eq. \eqref{eq:activity_lin} can also be more conveniently expressed in its logarithmic form:
\begin{equation}
    \ln a_{c} = \ln K_c(T) + \sum_{j \in \mathcal E} \nu_{cj} \ln n_j \ , \qquad\qquad c \in \mathcal C \ .
    \label{eq:activity_log}
\end{equation}
In the following, we will use this logarithmic form of the activity equation unless stated otherwise.

\subsubsection{Coupling of gas phase and condensates}

When condensates are taken into account, the element conservation equation is extended by a \textit{fictitious} number density of condensed molecules $n_c$:
\begin{equation}
    N_j = n_j + \sum_{i \in \mathcal S \setminus \mathcal E} \nu_{ij} n_i + \sum_{c \in \mathcal C_\text{S}} \nu_{cj} n_c \ , \qquad\qquad j \in \mathcal E \ ,
    \label{eq:element_conservation}
\end{equation}
where in analogy to Eq. \eqref{eq:def_nj} $N_j$ is defined as 
\begin{equation}
    N_j := \hat \epsilon_j n_\mathrm{\left\langle g\right\rangle} + \hat \epsilon_j n_\mathrm{\left\langle c\right\rangle}
\end{equation}
with
\begin{equation}
    n_\mathrm{\left\langle c\right\rangle} := \sum_{j \in \mathcal E} \sum_{c \in \mathcal C_S} \nu_{cj} n_c \ .
\end{equation}

This fictitious number density $n_c$ should not be confused with an actual dust particle number density. On the contrary, the $n_c$ are merely used in Eq. \eqref{eq:element_conservation} to keep track of the elements bound in the condensed material. It is also important to note that the fictitious number densities of the condensates do not contribute to the pressure of the gas phase in Eq. \eqref{eq:gas_pressure}.

An alternative way to describe the depletion of elements from the gas phase by condensates is the degree of condensation \citep[see][for example]{Gail2013pccd.book.....G}, given by:
\begin{equation}
    f_j = \frac{1}{N_j} \sum_{c \in \mathcal C_\text{S}} \nu_{cj} n_c \ , \qquad\qquad j \in \mathcal E\ .
    \label{eq:degree_of_condensation}
\end{equation}
Using this degree of condensation, the effective element abundance $\phi_j$ in the gas phase can be expressed via:
\begin{equation}
    \phi_j = \left(1 - f_j \right) \epsilon_j \ , \qquad\qquad j \in \mathcal E\ ,
    \label{eq:effective_element_abundances}
\end{equation}
where instead of $\epsilon_j$, $\phi_j$ is then used to calculate the normalised element abundances $\hat \epsilon_j$.

With the element abundances without considering condensation, $\hat \epsilon_j(f_j=0)$, and the total number density of atomic nuclei $n_\mathrm{\left\langle g\right\rangle}$, we can also define the largest possible fictitious number density $n_{c,\mathrm{max}}$ of a condensate:
\begin{equation}
\label{eq:cond_max_density}
   n_{c,\mathrm{max}} = \min_{\substack{j\in\mathcal E \\ \nu_{cj} > 0 }} \left\{ \frac{n_\mathrm{\left\langle g\right\rangle} \hat \epsilon_j(f_j=0)}{\nu_{cj}} \right\} \ , \qquad\qquad c \in \mathcal C\ ,
\end{equation}
assuming that the least common element in a condensate is completely condensed.

\subsubsection{Stability criterion for condensates}
\label{eq:stability_criterion}

The activity equation \eqref{eq:activity_log} provides a stability criterion for condensates. For those that exist in equilibrium with the gas phase, the activity has to be $a_c = 1$ and, thus,
\begin{equation}
    0 = \ln a_c =  \ln K_c(T) + \sum_{j \in \mathcal E} \nu_{cj} \ln n_j  \ , \qquad\qquad c \in \mathcal C_s \ .
     \label{eq:activity_eq}
\end{equation}
All other (unstable) condensates have an activity of $a_c < 1$, yielding:
\begin{equation}
    0 > \ln a_c = \ln K_c(T) + \sum_{j \in \mathcal E} \nu_{cj} \ln n_j  \ , \qquad\qquad c \in \mathcal C \setminus \mathcal C_s \ .
\end{equation}

In chemical equilibrium, therefore, only the following two different situations are permitted for all condensates:
\begin{itemize}
    \item stable condensates, $c \in \mathcal C_s$ \ : $\ln a_c = 0$, $n_c > 0$
    \item unstable condensates, $c \in \mathcal C \setminus \mathcal C_s$ \ : $\ln a_c < 0$, $n_c = 0$ \ .
\end{itemize}

\subsection{The phase rule}
\label{sec:gibbs_phase_rule}

Equation \eqref{eq:activity_eq} can be seen as constraints for the densities $n_j$ of the elements $\mathcal E_s$ that are contained in stable condensates. Since the gas phase contains $\lvert \mathcal E \rvert$ unique elements, there is a maximum number of $\lvert \mathcal E \rvert$ linearly-independent equations \eqref{eq:activity_eq}. This a priori limits the number of stable condensates to $\lvert \mathcal C_\text{s} \rvert \mathcal \leq \lvert \mathcal E \rvert$, which is usually a much smaller number than the total number of condensates $\lvert \mathcal C \rvert$ considered in a chemical system. 

As described in Sect. \ref{sec:gas_phase}, the $n_j$ are the result of a system of non-linear equations involving the law of mass action for the gas phase species, the element conservation equations, as well as the ideal gas law. In a case that includes stable condensates, however, some $n_j$ are also fixed by the solution of the Eq. \eqref{eq:activity_eq}. Consequently, they are therefore no longer free variables for the gas phase itself.

Thus, if all elements in the system are contained in stable condensates at the same time, their densities $n_j$ in the gas phase are all determined by Eqs. \eqref{eq:activity_eq}. The densities of all molecules are then directly given by their corresponding law of mass action with the fixed $n_j$ following Eq. \eqref{eq:mass_action_molecules}. This, however, creates the problem that the resulting $n_i$ and $n_j$ likely do not yield the correct pressure of the gas phase in Eq. \eqref{eq:gas_pressure} anymore. 

Therefore, at least one incondensable element in the gas phase is needed, whose $n_j$ is determined by the correct value of $p_{\mathrm g}$. Consequently, the total number of stable condensates is effectively limited to $\lvert \mathcal C_\text{s} \rvert \mathcal \leq \lvert \mathcal E \rvert - 1$. 

This is an extension of the famous, so-called Gibbs' phase rule \citep{Gib76,Gib78,Den55}. While the phase rule limits the number of stable condensates, it unfortunately does not directly provide information on the condensates contained in the set $\mathcal C_\text{s}$, allowing for a potentially large number of possible combinations.

\subsection{Solution strategies}

One fundamental difference between the law of mass action for molecules (Eq. \eqref{eq:mass_action_molecules}) in comparison to its formulation for condensates in Eq. \eqref{eq:activity_lin} is that the former directly yields the number density of a molecule if the $n_j$ are known, while the latter provides only a stability criterion. The number densities $n_c$ appear only in the element conservation equation \eqref{eq:element_conservation} and the degree of condensation \eqref{eq:degree_of_condensation}. Thus, there is no direct way of obtaining $n_c$.

This non-linear problem can be solved in a variety of different ways, two of which are briefly summarised in the following. For a broader overview of other methods, we refer the reader to \citet{vanZeggeren1970}.  We note, however, that the two schemes discussed next both assume that the set of stable condensates $\mathcal C_s$ is already known when the non-linear system is solved.

\subsubsection{Iterative solution with Newton's method}

One possible solution for the coupled problem is to use an iteration based on Newton's method. This, for example, is done in the \textsc{GGChem} code by \citet{Woitke2018A&A...614A...1W}. Their model uses the effective element abundances $\hat\epsilon_j$ and the fictitious condensate number densities $n_c$ as free variables. 

Using Eq. \eqref{eq:activity_eq} and the corresponding conservation equations for a considered set of stable condensates, \textsc{GGChem} assembles a linear system of equations to solve for the changes in the number densities $n_c$ for given changes in the $\hat\epsilon_j$. This is coupled to the gas phase calculations with $\hat\epsilon_j$ in a Newton's method scheme until $a_c=1$ is achieved for all considered condensates. In this case, the Jacobian matrix is not given analytically but is evaluated numerically by repeatedly calling the gas-phase-only calculations.

A Newton's method is also used in the so-called NASA-method approach \citep{Hoff1951}\footnote{The so-called NASA method by \citet{Hoff1951} should not be confused with the \textsc{NASA-CEA} code \citep{McBride96}. Both are based on very different computational methods: the former uses the law-of-mass action approach, while the latter is based on the minimisation of Gibbs free energy.}. In contrast to  \textsc{GGChem}, however, the NASA method solves the gas  and the condensed phase in a combined Newton's method rather than separating the compution into two distinct parts.

\subsubsection{Direct solution}

An alternative way is used in, for example, the \textsc{Condor} code by \citet{Lodders1993E&PSL.117..125L}. Here, Eq. \eqref{eq:activity_eq} for all considered, stable condensates is solved directly for the element densities $n_j$. Using these $n_j$ in the gas phase calculations and comparing the results with those of the condensate-free case allows to obtain the degree of condensation $f_j$ for all elements and, subsequently, the number densities $n_c$. No separate Newton's method is required here since the requirement of $a_c = 1$ is automatically satisfied by directly solving Eq. \eqref{eq:activity_eq} and keeping the $n_j$ fixed in the gas phase calculations.\\

\subsubsection{Deriving the set of stable condensates}

The two computational approaches discussed in the previous subsections assume that the set of stable condensates is already known. Usually, however, this is rarely the case and the solution of this problem therefore has to be embedded in some form of iterative scheme.

A typical calculation would, thus, start with determining the gas phase composition in the absence of any condensates. This provides initial guesses for the $n_j$. Using these initial solutions, one can test for the potential presence of stable condensates by calculating their activities. If all $a_c$ are smaller than unity, then no condensate can be present and the calculation is finished. 

Else, if some condensates have activities larger than unity, one then has to find a solution in terms of the element number density $n_j$ and the fictitious density $n_c$, such that the activities of all stable condensates are unity and the element conservation is fulfilled. The solution also has to agree with the phase rule and potentially stable condensates considered in the solution have to yield a linearly independent system of equations. 

Especially at low temperatures many different condensates can initially have activities $a_c > 1$ and, thus, are potential members of the set $\mathcal C_s$. This problem is commonly solved by constantly adding and removing condensates from $\mathcal C_s$, trying to solve the non-linear system with the methods described above \citep[e.g.][]{Gordon1994, Woitke2018A&A...614A...1W}. This iteration has to continue until all condensates in $\mathcal C_s$ have an activity of $a_c = 1$, while all others have $a_c < 1$. 

During the course of the iteration it might become necessary to remove condensates again from the set $\mathcal C_s$, if, for example, they yield negative densities $n_c$. Considering, however, that many condensates compete for the same elements it is not always immediately obvious, which condensates need to be removed to stabilise the system.

Given the large number of possible combinations of stable condensates at low temperatures, finding the final set $\mathcal C_s$ is, therefore, extremely challenging without any prior knowledge. One possibility is to start at high temperatures and slowly cool down the system, using the results of the higher temperatures as initial guesses for the lower temperature calculations \citep[e.g.][]{Burrows1999ApJ...512..843B}. \textsc{GGChem}, for example, additionally creates a lookup table of the effective element abundances as a function of pressure and temperature that allows to obtain the results in later calculations more easily. This lookup table, however, obviously depends directly on the individual element abundances and needs to be created for every set of metallicity values.

\section{Condensate treatment in FASTCHEM}
\label{sec:fastchem_cond}

As illustrated in the previous section, solving the combined condensate-gas phase problem is challenging, especially under low temperature conditions when many different condensates could potentially exist. The solution to this problem is made complicated by the fact that initially the final set of stable condensates $\mathcal C_s$ is not known. The phase rule and the requirement that the equations \eqref{eq:activity_eq} for all considered condensates need to be linearly independent offers some guidance. However, even with this supplementary information, an iterative approach of adding and removing condensates as described above does often not converge properly without further a priori information or good initial values for e.g. Newton's method \citep{Burrows1999ApJ...512..843B}.

The necessity of selecting an initial set $\mathcal C_s$ is caused by the fact that Eq. \eqref{eq:activity_eq} is only valid for stable condensates. Consequently, it is not possible to include all potential condensates at once in the system to derive the set $\mathcal C_s$ by solving the combined problem as evidently some of them might require a solution with $a_c < 1$ in case they are not stable.

In this new \fc version, we choose an alternative approach to this problem. This new approach is based on the basic ideas presented by \citet{Kulik2013} and \citet{Leal2016AdWR...96..405L}, the latter of which can be seen as a variant of the NASA method \citep{Hoff1951, vanZeggeren1970}. In the following, we adapt their equations and numerical methods to the general computational framework used in \fc.

\subsection{Modified condensate equations}

The most important modification of the standard approach is to adapt the activity equation in chemical equilibrium (Eq. \eqref{eq:activity_eq}), so that it is valid for \textit{all} condensate species $\mathcal C$ instead of only for species in the (initially unknown) set $\mathcal C_s$. This is achieved by introducing a correction factor $\lambda_c \geq 0$, such that
\begin{equation}
    0 = \ln a_c + \lambda_c =  \ln K_c(T) + \lambda_c + \sum_{j \in \mathcal E} \nu_{cj} \ln n_j  \ , \qquad c \in \mathcal C \ .
     \label{eq:activity_new}
\end{equation}

Obviously, for stable condensates ($c \in \mathcal C_s$), the correction factor is $\lambda_c = 0$, while for the unstable ones ($c \in \mathcal C \setminus C_s$), we require $\lambda_c = - \ln a_c$ in equilibrium.
As discussed by \citet{vanZeggeren1970} and \citet{Leal2016AdWR...96..405L}, the new correction coefficients $\lambda_c$ are directly connected to the Lagrange multipliers of the Gibbs free energy minimisation approach \citep[e.g.][]{Whi57,Whi58}.

Since a new variable $\lambda_c$ is introduced, an additional equation is required to solve the combined system of equations. This auxiliary equation combines $\lambda_c$ with the condensate number density $n_c$:
\begin{equation}
    \lambda_c \, n_c = 0 \ .
    \label{eq:activity_corr}
\end{equation}
This is supplemented by the element conservation equation, Eq. \eqref{eq:element_conservation}, involving all condensates. 

The combined system ensures that the two possible states for equilibrium condensation are a natural outcome of the problem, i.e.
\begin{itemize}
    \item $\lambda_c = 0 \Rightarrow \ln a_c = 0$, $n_c > 0 \ , \qquad\qquad c \in \mathcal C_s$
    \item $\lambda_c > 0 \Rightarrow \ln a_c < 0$, $n_c = 0 \ , \qquad\qquad c \in \mathcal C \setminus \mathcal C_s$ \ .
\end{itemize}
This approach, however, comes at the price of a larger system of equations, i.e. its dimension is in general $\left|\mathcal{E}\right|+2\left|\mathcal{C}\right|$ instead of $\left|\mathcal{E}_s\right|+\left|\mathcal{C}_s\right|$. 

For a numerical treatment, Eq. \eqref{eq:activity_corr} needs to be adapted slightly. The constraint that either $n_c$ or $\lambda_c$ need to be exactly equal to zero is numerically too strong. As suggested by \citet{Leal2016AdWR...96..405L} the equation is, therefore, modified with a constant $\tau_c$, such that
\begin{equation}
    \lambda_c \, n_c = \tau_c \ , \qquad\qquad c \in \mathcal C \ .
    \label{eq:activity_corr_mod}
\end{equation}
The constant $\tau_c$ is a very small number,  chosen such that it does not impact the outcome of calculation to a large degree. For $\tau_c$ we use the expression
\begin{equation}
    \tau_c = \tau \min_{\substack{j\in\mathcal E \\ \nu_{cj} > 0 }} \left\{ N_j \right\} \ , \qquad\qquad c \in \mathcal C \ ,
\end{equation}
where $\tau$ is a small number with a default value of $10^{-15}$. The quantity $\tau_c$ effectively controls the minimum non-zero number density of unstable condensates. The use of $\tau_c$ also implies that for stable condensates, $\ln a_c$ is not exactly zero but has a very small value of the order of $\tau$.

Since both $\lambda_c$ and $n_c$ can become very small numbers, we use the logarithmic form of Eq. \eqref{eq:activity_corr_mod}:
\begin{equation}
    \ln \lambda_c + \ln n_c - \ln \tau_c = 0 \ , \qquad\qquad c \in \mathcal C\ .
\end{equation}

In summary, we have to solve the following new, non-linear system of equations:
\begin{align}
    \label{eq:system_eq1}
    f_c^a &:= \ln K_c(T) + \lambda_c + \sum_{j \in \mathcal E} \nu_{cj} \ln n_j = 0 \ ,  \qquad &c \in \mathcal C\\
    \label{eq:system_eq2}
    f_c^\lambda &:= \ln \lambda_c + \ln n_c - \ln \tau_c = 0 \ ,  &c \in \mathcal C\\
    \label{eq:system_eq3}
    f_j^e &:= N_j - n_j - \sum_{i \in \mathcal S \setminus \mathcal E} \nu_{ij} n_i - \sum_{c \in \mathcal C} \nu_{cj} n_c = 0 \ ,     &j \in \mathcal E
\end{align}
for the unknown quantities $n_j$, $n_c$, and $\lambda_c$. Additionally, we require that all of these quantities have values larger than zero.

We note that the phase rule is never used explicitly in the construction of the system. Yet, the solution of the system of equations and, thus, the final set of stable condensates will automatically satisfy the phase rule. Furthermore, we also have not to pay any special attention to which condensates can be included to avoid linearly-dependent equations.

\subsection{Newton's method}

The combined system Eqs. \eqref{eq:system_eq1} - \eqref{eq:system_eq3} introduced in the last subsection is solved by using Newton's method with an analytic Jacobian matrix. Instead of solving for the variables $n_j$, $n_c$, and $\lambda_c$, we replace them by their logarithmic counterparts $\ln n_j$, $\ln n_c$, and $\ln \lambda_c$. This provides a better numerical stability and also guarantees that these quantities cannot become negative.

For a Newton step $k$ we perform first-order Taylor series expansions of the functions $f_c^\lambda$, $f_c^a$, and $f_j^e$ around the initial values $n_c^{(k)}$, $\lambda_c^{(k)}$, and $n_j^{(k)}$. For $f_c^\lambda$ this yields:
\begin{equation}
  \begin{split}
    0 & = f^\lambda_c\left((n_c^{(k)}, \lambda_c^{(k)}\right) + \frac{\partial f^\lambda_c}{\partial \ln n_c} \Delta \ln n_c + \frac{\partial f^\lambda_c}{\partial \ln \lambda_c} \Delta \ln \lambda_c\\
      & = \ln \lambda_c^{(k)} + \ln n_c^{(k)} - \ln \tau_c + \Delta \ln n_c + \Delta \ln \lambda_c \ ,
  \end{split}
\end{equation}
which can be re-written as
\begin{equation}
      b_c^\lambda =\Delta \ln n_c + \Delta \ln \lambda_c \ ,
      \label{eq:activity_corr_newton}
\end{equation}
with
\begin{equation}
      b_c^\lambda := \ln \tau_c - \ln \lambda_c^{(k)} - \ln n_c^{(k)}  \ .
      \label{eq:rhs_lambda}
\end{equation}

For the activity functions $f_c^a$ we perform the Taylor series expansion with respect to $\ln n_j$ and $\ln \lambda_c$:
\begin{equation}
    \begin{split}
    0 & = f^a_c\left(n_j^{(k)}, \lambda_c^{(k)}\right) + \sum_{j \in \mathcal E}\left(\frac{\partial f^a_c}{\partial \ln n_j} \Delta \ln n_j\right) + \frac{\partial f^a_c}{\partial \ln \lambda_c} \Delta \ln \lambda_c\\
      & =  \ln a_c^{(k)} + \lambda_c^{(k)} + \sum_{j \in \mathcal E} \nu_{cj} \Delta \ln n_j + \lambda_c^{(k)} \Delta \ln \lambda_c \ ,
    \end{split}
\end{equation}
where $\ln a_c^{(k)} = \ln a_c(\ln n_j^{(k)})$. This can also be written as
\begin{equation}
      b_c^a =  \sum_{j \in \mathcal E} \nu_{cj} \Delta \ln n_j + \lambda_c^{(k)} \Delta \ln \lambda_c \ ,
      \label{eq:activity_matrix_eq}
\end{equation}
with
\begin{equation}
      b_c^a :=  -\ln a_c^{(k)} - \lambda_c^{(k)} \ .
      \label{eq:rhs_activity}
\end{equation}

A similar expansion for the element conservation equations $f_j^e$ yields:
\begin{equation}
   \begin{split}
     0 & =  f^e_j\left(n_j^{(k)}, n_c^{(k)}\right) + \sum_{j' \in \mathcal E} \left( \frac{\partial f^e_j}{\partial \ln n_{j'}} \Delta \ln n_{j'} \right) + \sum_{c \in \mathcal C}\left(\frac{\partial f^e_j}{\partial \ln n_c} \Delta \ln n_c \right)\\
       & = N_j - n_j^{(k)} - \sum_{i \in \mathcal S \setminus \mathcal E} \nu_{ij} n_i^{(k)} - \sum_{c \in \mathcal C} \nu_{cj} n_c^{(k)} \\
       & \quad - \sum_{j'\in \mathcal E} \left( n_{j}^{(k)} \delta_{jj'} - \sum_{i \in \mathcal S \setminus \mathcal E} \nu_{ij} \nu_{ij'} n_i^{(k)} \right) \Delta \ln n_{j'} \\
       & \quad - \sum_{c \in \mathcal C} \nu_{cj} n_c^{(k)} \Delta \ln n_c \ .
   \end{split}
\end{equation}
Here we have used the law of mass action (Eq.~\eqref{eq:mass_action_molecules}) to analytically calculate the derivatives of the species densities $n_i$ with respect to those of the elements $n_j$ and to express the number densities $n_i^{(k)}$ as a function of the initial element densities $n_j^{(k)}$. This can be rearranged to:
\begin{equation}
  \begin{split}
    b_j^e =& \sum_{j'\in \mathcal E} \left( n_{j}^{(k)} \delta_{jj'} + \sum_{i \in \mathcal S \setminus \mathcal E} \nu_{ij} \nu_{ij'} n_i^{(k)} \right) \Delta \ln n_{j'} \\
    &+ \sum_{c \in \mathcal C} \nu_{cj} n_c^{(k)} \Delta \ln n_c \ ,
  \end{split}
  \label{eq:element_cons_matrix}
\end{equation}
with
\begin{equation}
    b_j^e := N_j - n_j^{(k)} - \sum_{i \in \mathcal S \setminus \mathcal E} \nu_{ij} n_i^{(k)} - \sum_{c \in \mathcal C} \nu_{cj} n_c^{(k)} \ .
    \label{eq:rhs_element_conserv}
\end{equation}

The final system of equations then has the general form
\begin{equation}
  \begin{pmatrix} 
  \boldsymbol{I} & \boldsymbol{I} & \boldsymbol{0}\\
  \boldsymbol{0} & \boldsymbol{\mathcal J_\mathcal{{\boldsymbol{\lambda}}}^a} & \boldsymbol{\mathcal J_\mathcal{E}^a}\\
  \boldsymbol{\mathcal J_\mathcal{C}^e} & \boldsymbol{0} & \boldsymbol{\mathcal J_\mathcal{E}^e}\\
  \end{pmatrix}
  \cdot
  \begin{pmatrix}
  \mathbf{\Delta \ln n_{\boldsymbol{\mathcal{C}}}}\\
  \mathbf{\Delta \ln \boldsymbol{\lambda}_{\boldsymbol{\mathcal{C}}}}\\
  \mathbf{\Delta \ln n_{\boldsymbol{\mathcal{E}}}}
  \end{pmatrix}
  = 
  \begin{pmatrix}
  \mathbf{b^{\boldsymbol{\lambda}}} \\
  \mathbf{b^a} \\
  \mathbf{b^e}
  \end{pmatrix} \ ,
  \label{eq:final_system_log_full}
\end{equation}
where $\boldsymbol{I}$ and $\boldsymbol{0}$ are the identity and zero matrix, respectively. The other components of the Jacobian matrix are given by the $\lvert \mathcal E \rvert \times \lvert \mathcal E \rvert$ matrix
\begin{equation}
  \left(\boldsymbol{\mathcal J_\mathcal{E}^e}\right)_{jj'} = n_j^{(k)} \delta_{jj'} + \sum_{i \in \mathcal S \setminus \mathcal E} \nu_{ij} \nu_{ij'} n_i^{(k)} \ ,
\end{equation}
the $\lvert \mathcal C \rvert \times \lvert \mathcal E \rvert$ matrix
\begin{equation}
  \left(\boldsymbol{\mathcal J_\mathcal{E}^a}\right)_{cj} = \nu_{cj} \ ,
\end{equation}
the $\lvert \mathcal E \rvert \times \lvert \mathcal C \rvert$ matrix
\begin{equation}
  \left(\boldsymbol{\mathcal J_\mathcal{C}^e}\right)_{jc} = \nu_{cj} n_c^{(k)} \ ,
\end{equation}
as well as the $\lvert \mathcal C \rvert \times \lvert \mathcal C \rvert$ diagonal matrix
\begin{equation}
  \left(\boldsymbol{\mathcal J_\mathcal{{\boldsymbol{\lambda}}}^a}\right)_{cc} = \lambda_c^{(k)} \ .
\end{equation}
The components of the vector of unknowns are:
\begin{equation}
  \begin{split}
    \left(\mathbf{\Delta \ln n_{\boldsymbol{\mathcal{C}}}}\right)_c &= \Delta \ln n_c\\
    \left(\mathbf{\Delta \ln \boldsymbol{\lambda}_{\boldsymbol{\mathcal{C}}}}\right)_c &= \Delta \ln \lambda_c\\
    \left(\mathbf{\Delta \ln n_{\boldsymbol{\mathcal{E}}}}\right)_j &= \Delta \ln n_j \ ,
  \end{split}
\end{equation}
while those of the right-hand-side vector are given by Eqs. \eqref{eq:rhs_lambda}, \eqref{eq:rhs_activity} \& \eqref{eq:rhs_element_conserv}. The total number of unknowns for this system is $2 \, \lvert \mathcal C \rvert + \lvert \mathcal E \rvert$. Technically speaking, the system of equations is usually smaller than that because elements not forming any potentially stable condensate don't need to be included in the system. 

With the computed solution for the quantities $\Delta \ln n_j$, $\Delta \ln n_c$, and $\Delta \ln \lambda_c$, the corresponding quantities are then corrected following 
\begin{equation}
    \ln x^{(k+1)} = \ln x^{(k)} + \Delta \ln x \ ,
\end{equation}
where $x$ is either $n_j$, $n_c$, or $\lambda_c$. As mentioned earlier, the use of logarithmic quantities prevents these values from becoming negative or zero by construction.

\subsection{Reduction of the system of equations}

The system of equations \eqref{eq:final_system_log_full} can be become rather large if the number of potential condensates is very high, which typically happens at low temperatures. Its size, however, can be reduced significantly by removing both $\Delta \ln \lambda_c$ and $\Delta \ln n_c$ from the system of equations.

To remove $\Delta \ln \lambda_c$ we first solve Eq. \eqref{eq:activity_corr_newton} for $\Delta \ln \lambda_c$:
\begin{equation}
    \Delta \ln \lambda_c = \ln \tau_c - \lambda_c^{(k)} - \ln n_c^{(k)} - \Delta \ln n_c \ .
    \label{eq:def_delta_lambda}
\end{equation}
In Eq. \eqref{eq:activity_matrix_eq} we can now replace the quantity $\ln \Delta \lambda_c$, which yields
\begin{equation}
    \hat{b}_c^a = \sum_{j \in \mathcal E} \nu_{cj} \Delta \ln n_j - \lambda_c^{(k)} \Delta \ln n_c  \ ,
    \label{eq:activity_matrix}
\end{equation}
with
\begin{equation}
    \hat{b}_c^a := - \ln a_c^{(k)} - \lambda_c^{(k)} \left(1 + \ln \tau_c - \ln \lambda_c^{(k)} - \ln n_c^{(k)} \right) \ .
    \label{eq:rhs_activity_alt}
\end{equation}

This reduces the linear system of equations to the form
\begin{equation}
  \begin{pmatrix} 
  -\boldsymbol{\mathcal J_\mathcal{{\boldsymbol{\lambda}}}^a} & \boldsymbol{\mathcal J_\mathcal{E}^a}\\
  \boldsymbol{\mathcal J_\mathcal{C}^e} & \boldsymbol{\mathcal J_\mathcal{E}^e}\\
  \end{pmatrix}
  \cdot
  \begin{pmatrix}
  \mathbf{\Delta \ln n_{\boldsymbol{\mathcal{C}}}}\\
  \mathbf{\Delta \ln n_{\boldsymbol{\mathcal{E}}}}
  \end{pmatrix}
  = 
  \begin{pmatrix}
  \mathbf{\hat{b}^a} \\
  \mathbf{b^e}
  \end{pmatrix} 
  \label{eq:final_system_log_small}
\end{equation}
for the $\lvert \mathcal C \rvert + \lvert \mathcal E \rvert$ unknowns $\Delta \ln n_c$ and $\Delta \ln n_j$. With the $\Delta \ln n_c$ from the solution of this system, we can then calculate the corresponding corrections $\Delta \ln \lambda_c$ from Eq. \eqref{eq:def_delta_lambda}.

The same approach can be used for the unknowns $\Delta \ln n_c$ \citep{Leal2016AdWR...96..405L}. Using Eq. \eqref{eq:activity_matrix}, we can derive an equation for $\Delta \ln n_c$:
\begin{equation}
  \begin{split}
    \Delta \ln n_c =& \frac{1}{\lambda_c^{(k)}} \sum_{j \in \mathcal E} \nu_{cj} \Delta \ln n_j - \frac{b_c^a}{\lambda_c^{(k)}}\\
                   =& \frac{1}{\lambda_c^{(k)}} \sum_{j \in \mathcal E} \nu_{cj} \Delta \ln n_j \\
                     &+ \frac{\ln a_c^{(k)}}{\lambda_c^{(k)}} +\ln{\tau_c} - \ln{n_c^{(k)}} -\ln{\lambda_c^{(k)}} + 1
  \end{split}
  \label{eq:def_delta_nc}
\end{equation}
and use it to eliminate $\Delta \ln n_c$ in Eq. \eqref{eq:element_cons_matrix}. This yields a system of equations where the $\Delta \ln n_j$ are the only unknowns:
\begin{equation}
    \hat{b}_j^e = \sum_{j' \in \mathcal E} \left( \left(\boldsymbol{\mathcal J_\mathcal{E}^e}\right)_{jj'} + \sum_{c \in \mathcal C}  \nu_{cj} \nu_{cj'} \frac{n_c^{(k)}}{\lambda_c^{(k)}} \right) \Delta \ln n_{j'} \ ,
    \label{eq:jacobian_reduced}
\end{equation}
with the right-hand-side vector components
\begin{equation}
  \hat{b}_j^e := b_j^e - \sum_{c \in \mathcal C} \nu_{cj} n_c^{(k)} \left(\frac{\ln a_c^{(k)}}{\lambda_c^{(k)}} +\ln{\tau_c} - \ln{n_c^{(k)}} -\ln{\lambda_c^{(k)}} + 1\right) \ .
\end{equation}
With the resulting $\Delta \ln n_j$ we can calculate the corrections $\Delta \ln n_c$ from Eq. \eqref{eq:def_delta_nc} and subsequently again $\Delta \ln \lambda_c$ from Eq. \eqref{eq:def_delta_lambda}.

Instead of the initial $\lvert \mathcal E \rvert + 2 \lvert \mathcal C \rvert$ dimensions, the system of equations is now reduced to just $\lvert \mathcal E \rvert$ equations for the elements. Given that usually $\lvert \mathcal E \rvert \ll \lvert \mathcal C \rvert$, this provides a huge increase in computational performance.

Unfortunately, using this replacement for all condensates can yield a numerically unstable system. The reduced Jacobian matrix in Eq. \eqref{eq:jacobian_reduced} has terms of the form $n_c/\lambda_c$. For stable condensates, $\lambda_c$ approaches values of close to zero. In fact, with the original equation \eqref{eq:activity_corr} we would effectively perform a division by zero here in case of stable condensates with $n_c \gg 1$. This generates very large, off-diagonal entries in the Jacobian, which can potentially make the matrix essentially numerically singular.

For unstable condensates we have $n_c \approx 0$ and $\lambda_c \approx -\ln a_c$. In such a case, the problematic terms vanish from the Jacobian and the matrix remains well-conditioned. Thus, while stable condensates have to be included in the Jacobian matrix for numerical stability, unstable ones can be safely removed. The system of equations would, therefore, have a dimension of $\lvert \mathcal C_s \rvert + \lvert \mathcal E \rvert$ once all unstable condensates have been identified and accordingly removed from the Jacobian matrix.

\subsection{Numerical solution}

To allow for more flexibility in the numerical calculations, \fcc contains all three different system of equations introduced so far. Before any of these systems of equations can be solved, the corresponding Jacobian matrix requires some rescaling to ensure numerical stability. The different matrix components of the Jacobian imply that entries can potentially differ by many orders of magnitudes based on the current values of $n_c$, $\lambda_c$, or $n_i$. This can result in an unstable numerical behaviour. Before solving the system, each row of the matrix is therefore scaled by the inverse of its largest entry. Mathematically, this corresponds to the multiplication of the Jacobian with a non-singular, diagonal scaling matrix \citep[cf.][]{DeuflhardNewton} and usually leads to a well-conditioned Jacobian matrix.

We solve the system of equations directly by using an LU decomposition with partial pivoting from the Eigen library\footnote{\url{https://eigen.tuxfamily.org}}. This method provides a fast and stable way to obtain the solution. In principle, we could also exploit the fact that the Jacobian has a very sparse structure, since it contains more zero entries than non-zero ones. Thus, we could use iterative methods that have been especially designed for such systems with a sparse coefficient matrix. However, the overall dimension of our system is normally still quite small and these iterative methods, therefore, tend not to perform better than the direct solution via the LU decomposition. 

In some cases, the system of equations can still become ill-conditioned which makes a numerical solution challenging. Especially the Jacobians for the reduced systems \eqref{eq:final_system_log_full} and \eqref{eq:jacobian_reduced} are prone to become numerically singular. If, through the course of the Newton iterations, both $n_c$ and $\lambda_c$ for certain condensates become very small, the corresponding coupling terms in the submatrices $\boldsymbol{\mathcal J_\mathcal{C}^e}$ and $\boldsymbol{\mathcal J_\mathcal{{\boldsymbol{\lambda}}}^a}$ for these condensates are so small that they yield an almost singular matrix. This, of course, creates problems when using the LU decomposition with partial pivoting because this method a priori assumes that the matrix is invertible. Thus, the numerical solution for such an ill-conditioned system can in general not be trusted. 

The full system \eqref{eq:final_system_log_full} is much less affected by these issues, since there are still coupling terms remaining in the Jacobian that stabilise the numerical solution. However, in rare cases even the full Jacobian matrix can become ill-conditioned. 

The Eigen library also offers, for example, an LU decomposition with full pivoting. While this method is slower than the version with partial pivoting, it can detect cases with singular matrices. \fcc is able to use both solvers but by default employs the one with partial pivoting as it provides a much higher computational speed. The user can optionally choose the LU decomposition with full pivoting.

If \fcc uses the method with full pivoting and detects a non-invertible matrix, it will either use a singular value decomposition to solve the system of equations or perturbs the Jacobian to make it invertible \citep{DennisSchnabelNumericalMethods}. However, we note that both methods can lead to solutions that produce so minor corrections, such that the Jacobian matrix will remain non-invertible in the next iteration step. In that case, the iterative method in \fcc will fail to converge. Consequently, in such a situation it is usually advantageous to take a dampened Newton step into any random direction provided by the unstable LU decomposition. Most of the time this produces a well-conditioned matrix again in the next iteration step.

We also note that the full system of equations \eqref{eq:final_system_log_full} is much less affected by a potentially singular Jacobian matrix. Even in the case of very small values for $n_c$ and $\lambda_c$, there are still additional coupling terms left in the matrix to yield a stable matrix inversion. In fact, using the full system tends to require less iterations to converge than the reduced system of equations. However, the numerical cost per iteration step is much higher, which makes this approach overall often slower in terms of computational time. \fcc will switch to this method if the iteration with the reduced system does not converge. The user can also optionally force \fcc to always use the full Jacobian matrix in its calculations, which might help to overcome convergence issues.

\subsection{Coupling with the gas phase chemistry}

The condensation part of \fcc described above, needs to be iteratively coupled to the gas phase chemistry calculations. At the start of each coupled calculation, \fc will first determine the gas phase composition, neglecting any condensates. With that solution, the activities of all condensates are obtained. In case no activity is larger than unity, no stable condensates are possible and the calculation is finished. 

If \fc finds some species, with initial $a_c$ larger than one, it selects these condensates as potential candidates for the condensed phase and proceeds with the solution of the system described in the previous section. In principle, the mathematical description allows to include \textit{all} condensates available in the chemical model, even those with activities smaller than unity. Since, however, those are not stable by definition, only condensates that can potentially be present are included in the calculation to limit the computational cost of solving the system of equations.

Having determined the set stable condensates and their corresponding densities $n_c$, the depletion of the elements from the gas phase is evaluated using the degree of condensation from Eq. \eqref{eq:degree_of_condensation} and the effective element abundances $\phi_j, j\in\mathcal{E}_s$. In principle, one could now use these $f_j$ and $\phi_j$ values to re-calculate the gas phase for all elements. However, at low temperatures some of the $f_j$ can become numerically exactly unity, which yields an effective element abundance of $\phi_j = 0$. Consequently, this can create numerical issues in the gas phase calculations for these elements.

Instead, we fix the number densities of the elements in their atomic form, $n_j$, in the gas-phase calculations to the solution we obtain from the calculation of the condensed phase. As mentioned earlier, these $n_j$ are essentially fixed by the solution of the activity equations and, therefore, should actually not be considered anymore as free variables in the gas phase.

An obvious advantage of using the second approach is that the calculation of the gas phase is simplified, since some elements do not need to be calculated at all as their $n_j$ are already determined. This is especially helpful at lower temperatures when potentially many elements are taking part in condensation.

\fcc will iteratively perform the calculations of the condensed and gas phase chemistry until the coupled system is converged. Normally, the first condensation iteration takes the longest time since this is the step where the selection of the stable and unstable condensates is made. Later calls of the condensation system typically need much fewer Newton steps because the sets of stable and unstable condensates usually don't change significantly. After each calculation of the gas phase \fcc checks for additional, potential condensates that need to be added to the system. The calculation proceeds until all condensates fulfill their stability criterion from Sect. \eqref{eq:stability_criterion} and all elements are conserved. 

\begin{figure*}
	\resizebox{\columnwidth}{!}{\includegraphics{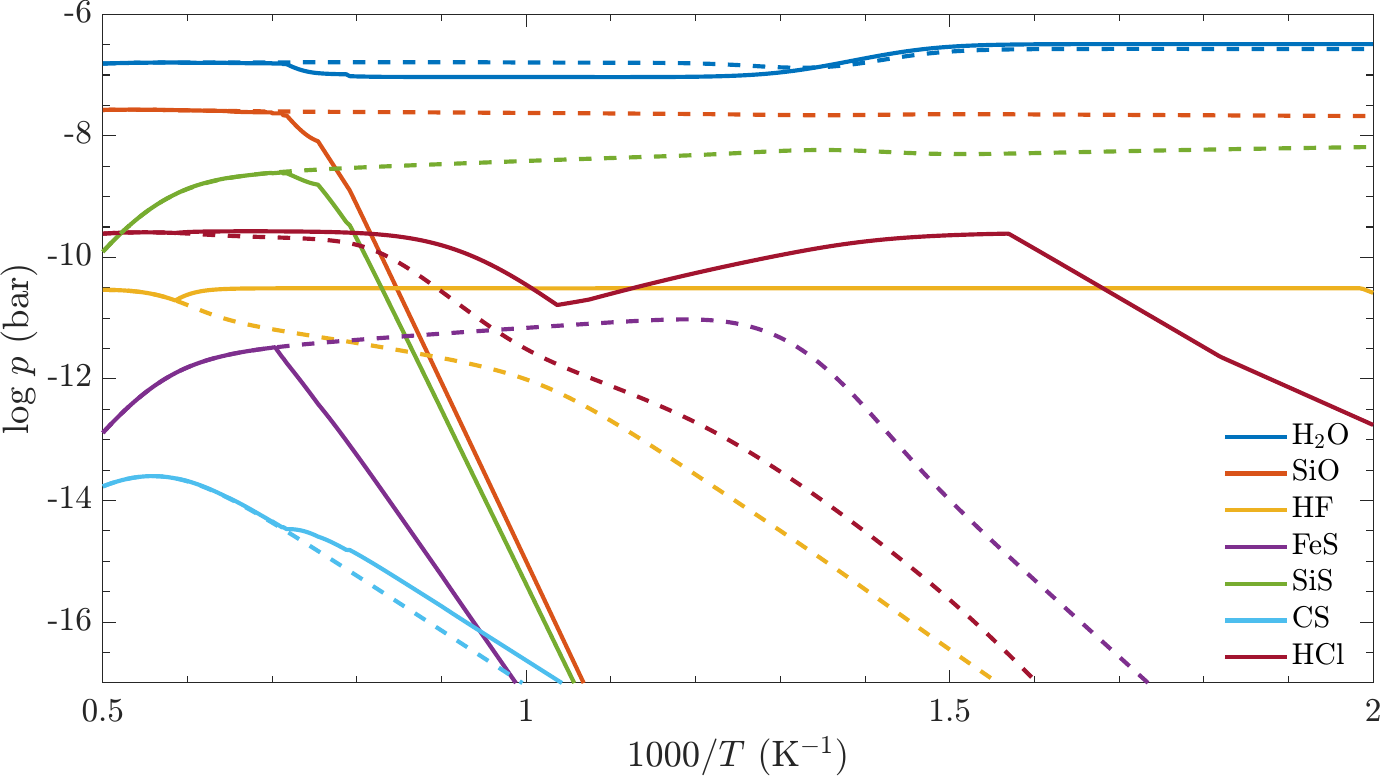}}	
	\resizebox{\columnwidth}{!}{\includegraphics{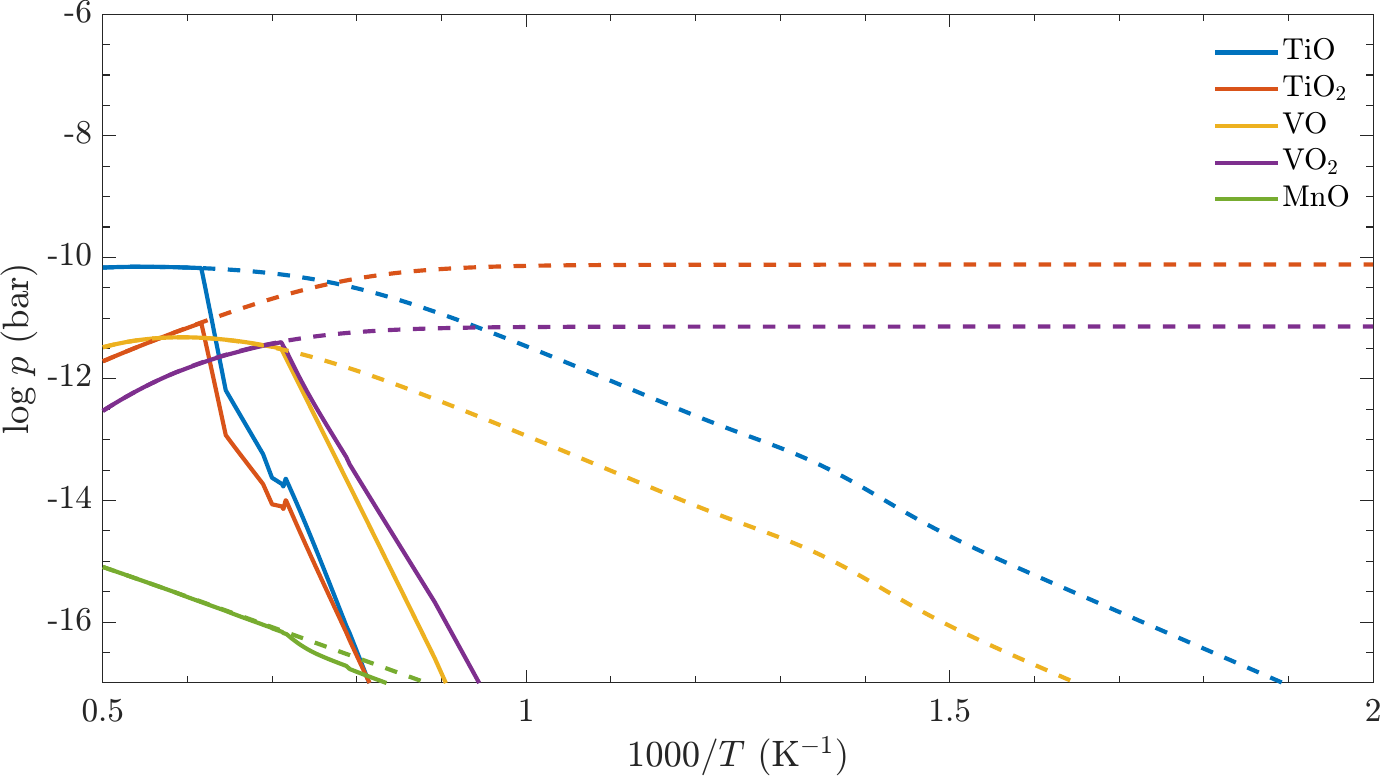}}\\
	\resizebox{\columnwidth}{!}{\includegraphics{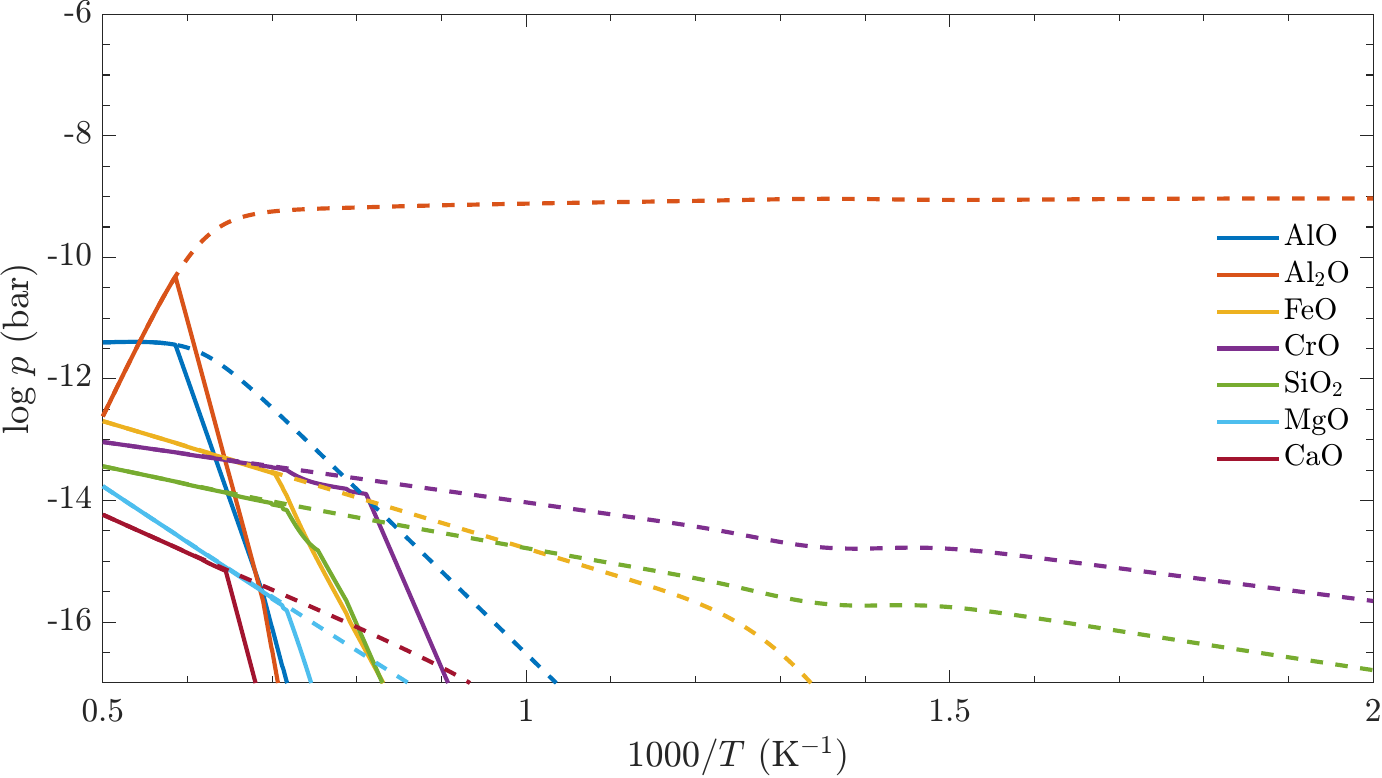}}	
	\resizebox{\columnwidth}{!}{\includegraphics{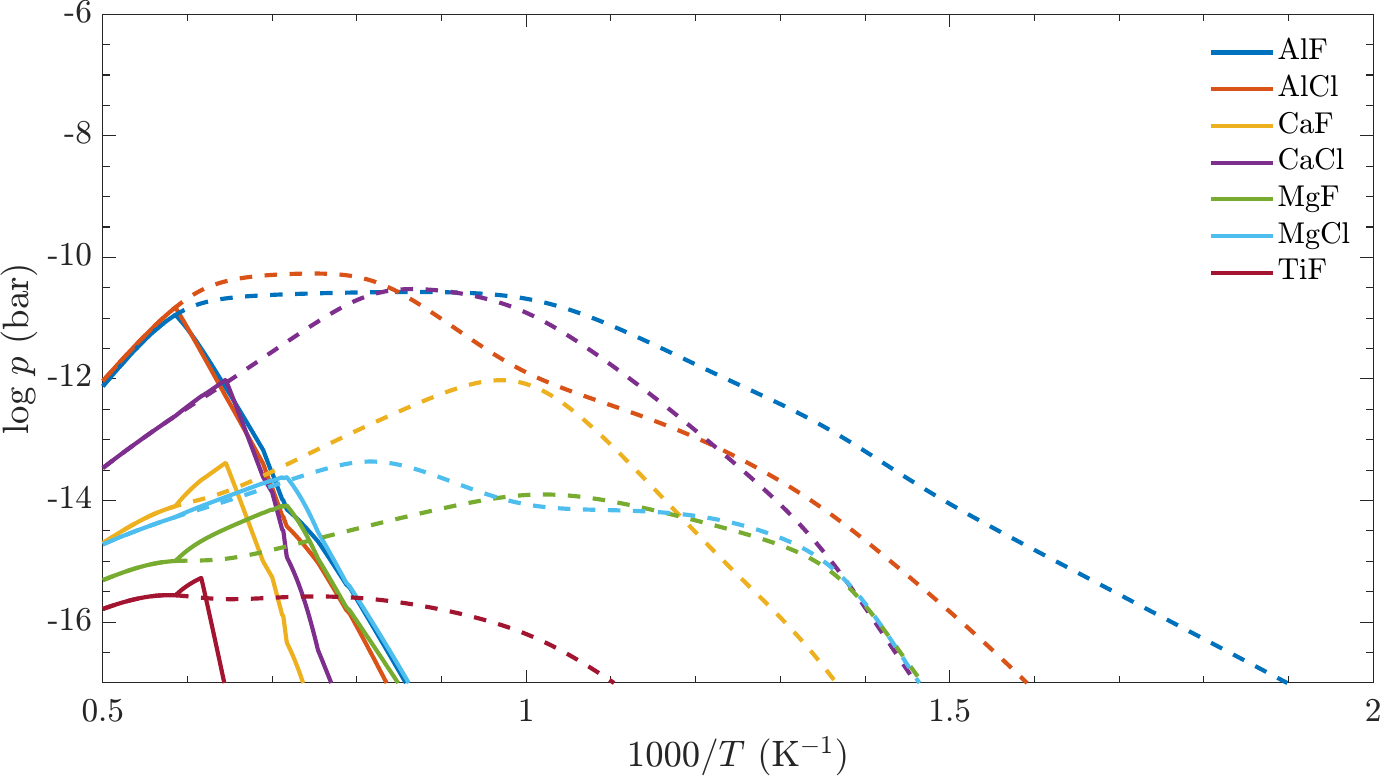}}\\	
	\resizebox{\columnwidth}{!}{\includegraphics{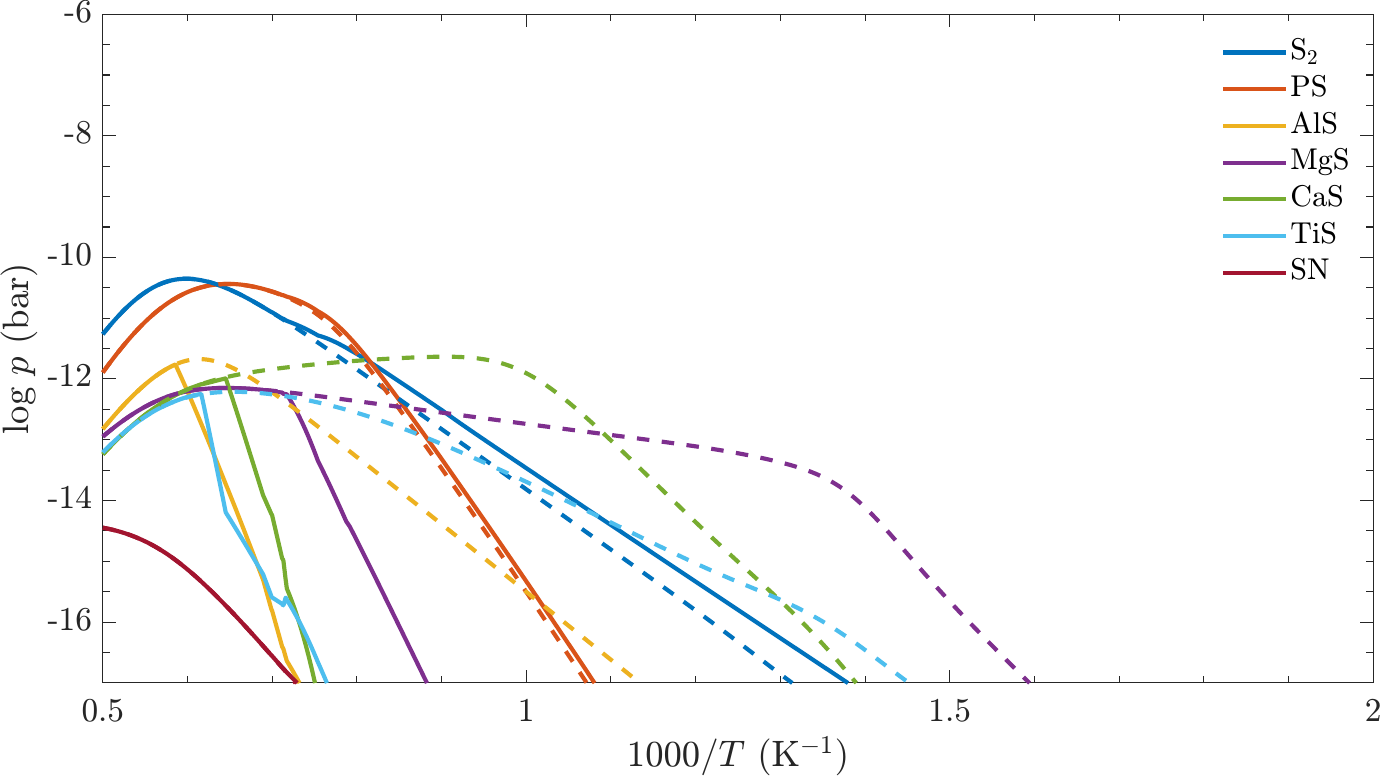}}	
	\resizebox{\columnwidth}{!}{\includegraphics{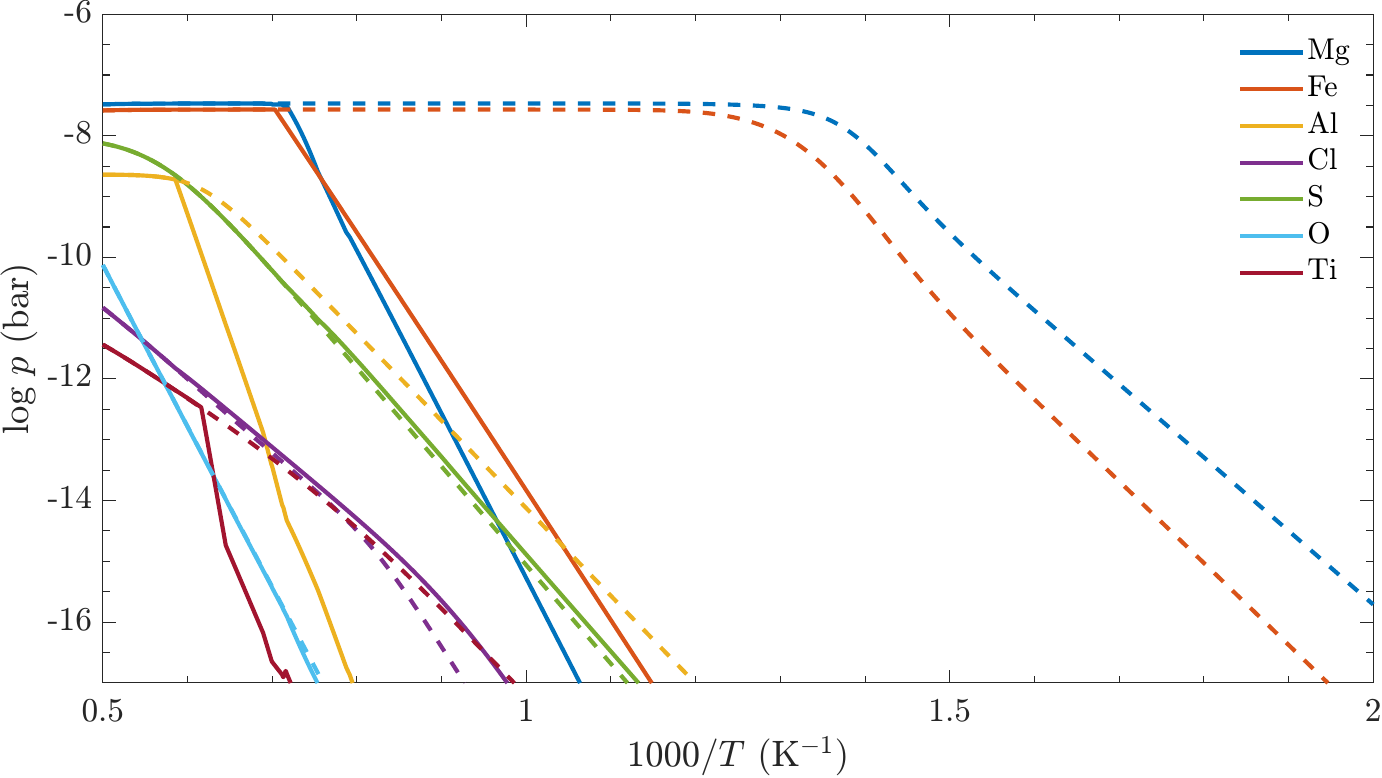}}	
	\caption{Comparison with equilibrium condensation calculations of \citet{Sharp1990ApJS...72..417S}. The plots show the partial pressures of gas-phase species as a function of temperature and roughly mirror the results shown by \citet{Sharp1990ApJS...72..417S} with some species moved to other panels for better visualisation. We note that that our calculations extend to lower temperatures than those presented by \citet{Sharp1990ApJS...72..417S}. Following the corresponding figures by \citet{Sharp1990ApJS...72..417S}, solid lines refer to the chemistry calculations with equilibrium condensation while dashed lines refer to those without.}
	\label{fig:sh_1}
\end{figure*}

\subsection{Rainout approximation}
\label{sec:rainout}

Besides the standard equilibrium condensation, \fcc also supports calculations using the so-called rainout approximation. In the former case, each temperature and pressure point is treated individually. However, as pointed out by, for example, \citet{Lodders2002Icar..155..393L} or \citet{Marley2013cctp.book..367M} this pure equilibrium condensation is not suitable for calculating the chemical composition of planetary atmospheres or brown dwarfs, because rainout affects the element distribution as function of altitude. 

As discussed by \citet{Lodders2002Icar..155..393L}, condensates in such high-gravity environments tend form directly from the gas phase (\textit{primary} condensates) that then settle into cloud layers. In a low-gravity case, such as a protoplanetary disk or a pre-stellar nebula, these primary condensates do not settle into distinct layers but remain dispersed in the gas phase. This allows for the formation of \textit{secondary} condensates via chemical reactions between the gas phase and (primary) condensates when the temperature decreases further. 

To simulate the rainout, \fcc  will first start the calculation at the lower boundary (highest pressure) for a given temperate-pressure profile $T(p)$ and work its way upwards towards lower pressures. If condensation is encountered at some pressure $p$, \fcc will solve the coupled condensation and gas phase system as previously described. The converged system then yields the effective element abundances $\phi_j$ of the condensing elements left in the gas phase (see Eq. \eqref{eq:effective_element_abundances}). Using these effective abundances, \fc will then change the actual element abundances $\epsilon_j$ to the computed $\phi_j$ at all pressures below the level where condensation occurred. This simulates the rainout of these elements into condensate layers and leads usually to a rapid decrease in the abundances of condensing elements in the upper parts of an atmosphere.

An example to illustrate the differences between equilibrium condensation and rainout is shown in Sect. \ref{sec:rainout_example} for an atmosphere of a brown dwarf.

\subsection{Code update and availability}

The numerical treatment for condensation and rainout described in the previous subsections has been added to the \fc code. It is available as open source on GitHub (\url{https://github.com/exoclime/FastChem}). Like the previous versions of \fc, the code is released under the GNU General Public License version 3 \citep{Gnu07}. This new \fc release has a version number of 3.0 and is referred to as \fcc. We note that for calculations not involving condensation, the results of \fcc are identical to those calculated with the previous version \fc 2.

As described by \citet{Stock2022MNRAS.517.4070S}, \fc is written mainly in object-oriented C++ but additionally offers a Python interface (\textsc{pyFastChem}) that allows it to be imported as a normal Python module. Some example Python scripts are included in the repository that showcase the use of \fcc for several different scenarios. In addition to the GitHub repository, \textsc{pyFastChem} is also available as a \texttt{PyPI} package that can easily be installed via \texttt{pip}. 

The Python interface of \fcc has been adapted to account for calculations involving condensates. New parameter options have also been included in the interface that allow the user to access special internal parameters. A detailed description of the C++ code and the Python interface is available in the manual, located in the GitHub repository.

Thermochemical data for condensates mostly based on the JANAF tables \citep{Chase1998} has been added to \fcc. In particular, the current version of \fcc includes about 290 solids and liquids for 27 elements more abundant than germanium for the solar element abundances by \citet{Asplund09}. Details on the condensate data can be found in Appendix \ref{sec:data_fits}. We aim to add more elements, gas phase species, and condensates in future releases of \fc.

\section{Test calculations}
\label{sec:model_tests}

In this section we test the new condensation approach in \fcc for three different scenarios of astrophysical interest. In the first case we replicate the equilibrium condensation calculations from \citet{Sharp1990ApJS...72..417S}. In the second scenario we evaluate the numerical stability of our new code by calculating the equilibrium condensation in a protoplanetary disk. Since the temperatures within the disk drop to values of roughly 10 K, this scenario presents a numerically very challenging case. In the last scenario we show the differences in the chemical composition between a standard equilibrium condensation calculation and the rainout approximation for the atmosphere of a brown dwarf. All results shown in the following can be easily reproduced by the Python scripts contained in the \fc repository.

\subsection{Comparison with Sharp \& Huebner (1990)}

The study by \citet{Sharp1990ApJS...72..417S} focuses on the chemical composition of the gas phase under the influence of condensation in a general astrophysical context. They adapted the SOLGASMIX code \citep{Eriksson1971, Besmann1977} to include the condensation of species for solar element abundances. In contrast to \fcc, SOLGASMIX used a Gibbs free energy minimisation approach to solve the equilibrium condensation problem. The study presents the resulting gas phase composition as a function of temperature for a fixed pressure of 0.5 bar. 

For this test calculation, we replicate the temperature and pressure conditions used by \citet{Sharp1990ApJS...72..417S}. The resulting abundances of selected gas phase species are shown in the plots of Fig. \ref{fig:sh_1}, which roughly mirror those presented by \citet{Sharp1990ApJS...72..417S}. We note, however, that we extended the temperature range of the chemistry calculations to even lower temperatures than those used by \citet{Sharp1990ApJS...72..417S} to illustrate the numerical capabilities of our new code.

Comparing our results with those of \citet{Sharp1990ApJS...72..417S} suggests a very good agreement in terms of the resulting molecular abundances. Minor changes can be explained by the different element abundances. \citet{Sharp1990ApJS...72..417S} employed element abundances based on \citet{Cameron1973SSRv...15..121C}, wheres \fcc uses the ones from \citet{Asplund09}. Other notable differences are the different sets of gas phase species and condensates as well as the underlying thermochemical data, the fitted equilibrium constants are based on.

\begin{figure}
	\resizebox{\columnwidth}{!}{\includegraphics{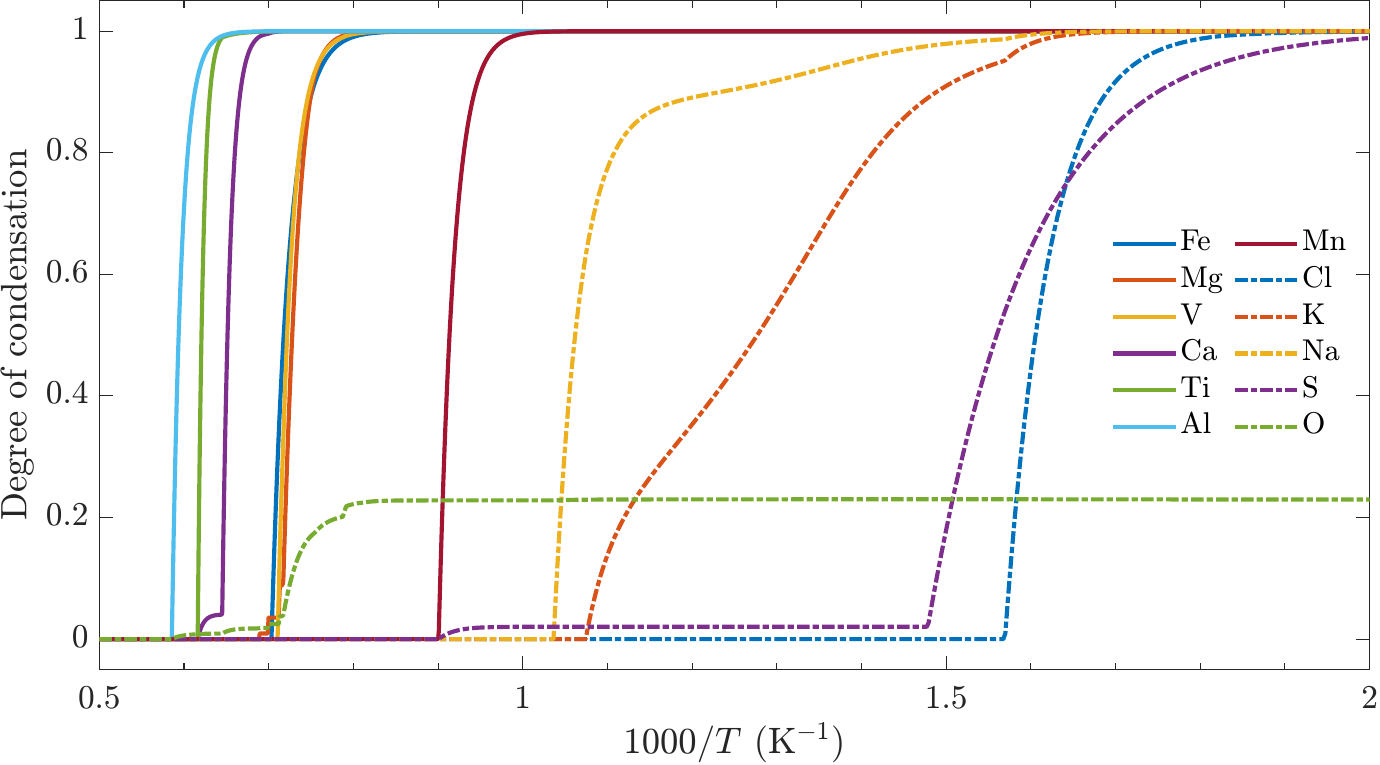}}
	\caption{Degree of condensation for selected elements for the \citet{Sharp1990ApJS...72..417S} scenario as a function of temperature and a gas pressure of $p_\mathrm{g} = 0.5$ bar.}
	\label{fig:sh_cond_degree}
\end{figure}

In addition to the gas phase, we also show the degree of condensation for several elements in Fig. \ref{fig:sh_cond_degree}. The results indicate that most species condense out completely below a certain temperature. Because oxygen is much more abundant than the other more refractory elements, only about 20\% of O is condensed out until the lowest considered temperature of 500 K. For a solar element abundance mixture, oxygen would condense out almost completely at even lower temperatures in the form of water ice.

\begin{figure}
	\resizebox{\columnwidth}{!}{\includegraphics{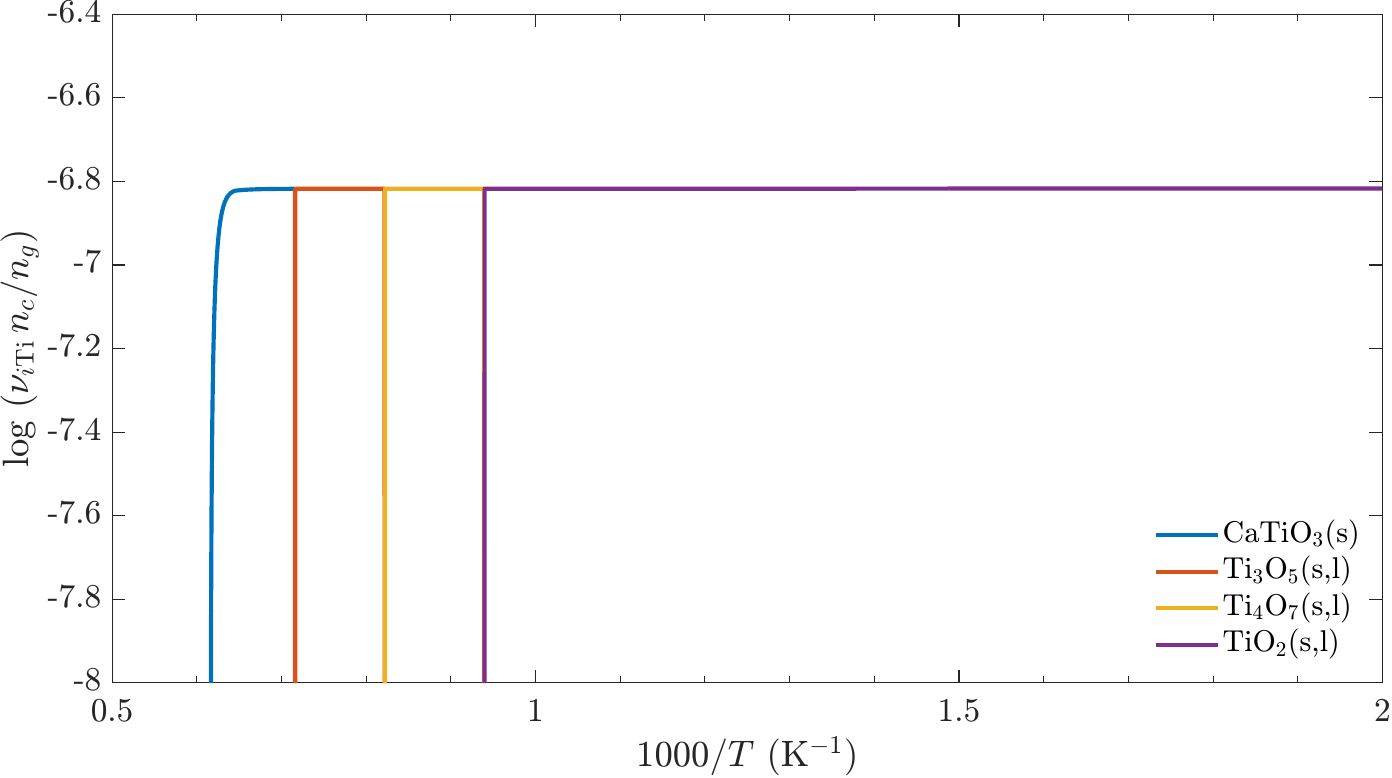}}\\
	\resizebox{\columnwidth}{!}{\includegraphics{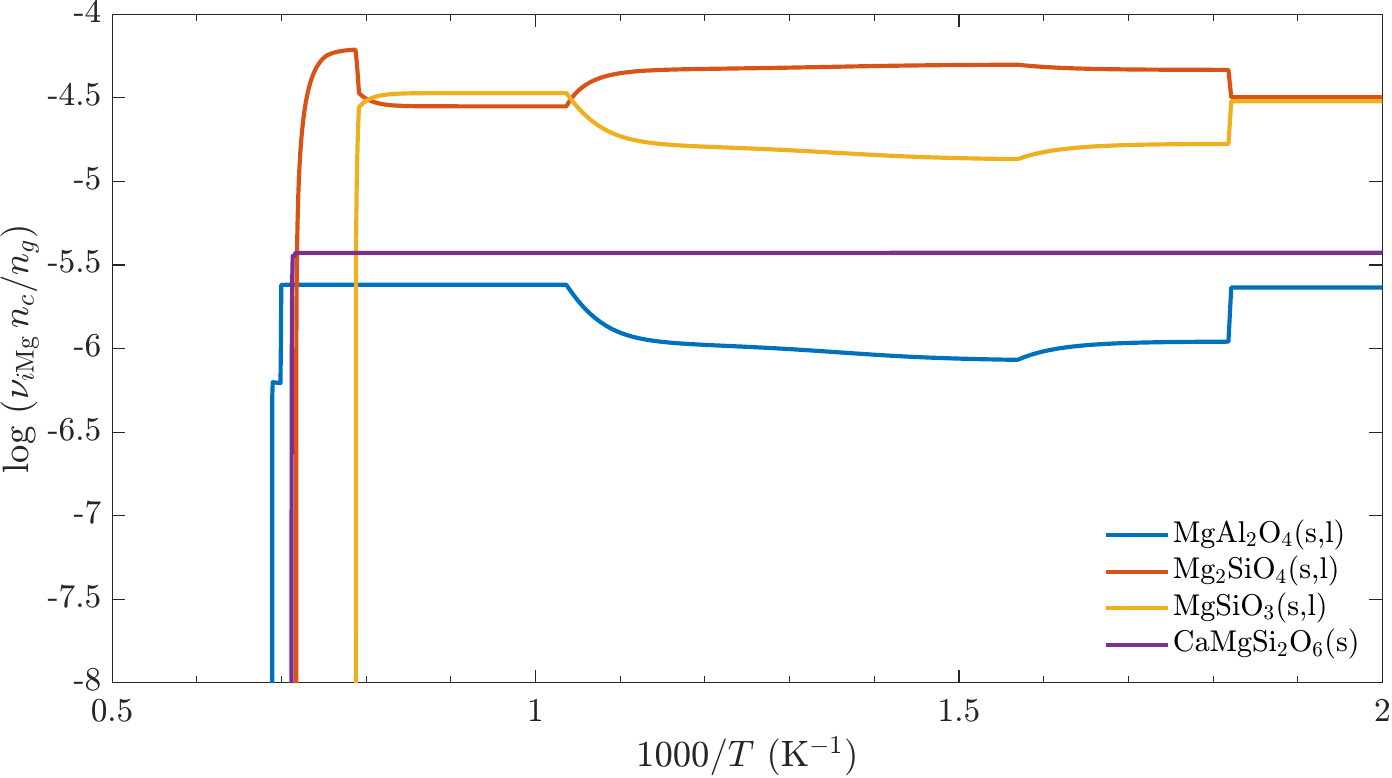}}
	\caption{Examples for condensation sequences of selected elements for the \citet{Sharp1990ApJS...72..417S} scenario. The upper panel shows titanium-bearing condensates, while the lower panel depicts the condensation sequence for magnesium.}
	\label{fig:sh_cond_seq}
\end{figure}

Examples for the condensation sequences of titanium and magnesium-bearing condensates are shown in Fig. \ref{fig:sh_cond_seq}. Titanium first condenses into the high-temperature condensate perovskite (\ch{CaTiO3(s)}). At lower temperatures, \ch{Ti} then switches to titanium oxide condensates, namely \ch{Ti3O5}, \ch{Ti4O7}, and \ch{TiO2}. The transitions between the different Ti condensates are very sharp, suggesting that titanium is only present in a single condensate at a time.

In contrast, magnesium can be found in several stable condensates simultaneously as suggested by the lower panel of Fig. \ref{fig:sh_cond_seq}. It first condenses into spinel (\ch{MgAl2O4}), followed shortly afterwards by diopside (\ch{CaMgSi2O6}) and forsterite (\ch{Mg2SiO4}). Lastly, magnesium also condenses into enstatite (\ch{MgSiO3}), which competes with forsterite for the most abundant magnesium-bearing condensate. In contrast to the titanium condensation sequence, none of the magnesium condensates actually disappear after they have formed, though their abundances can change as a function of temperature.

\subsection{Protoplanetary disk}
\label{ssec:proto_disk}

\begin{figure}
	\resizebox{\columnwidth}{!}{\includegraphics{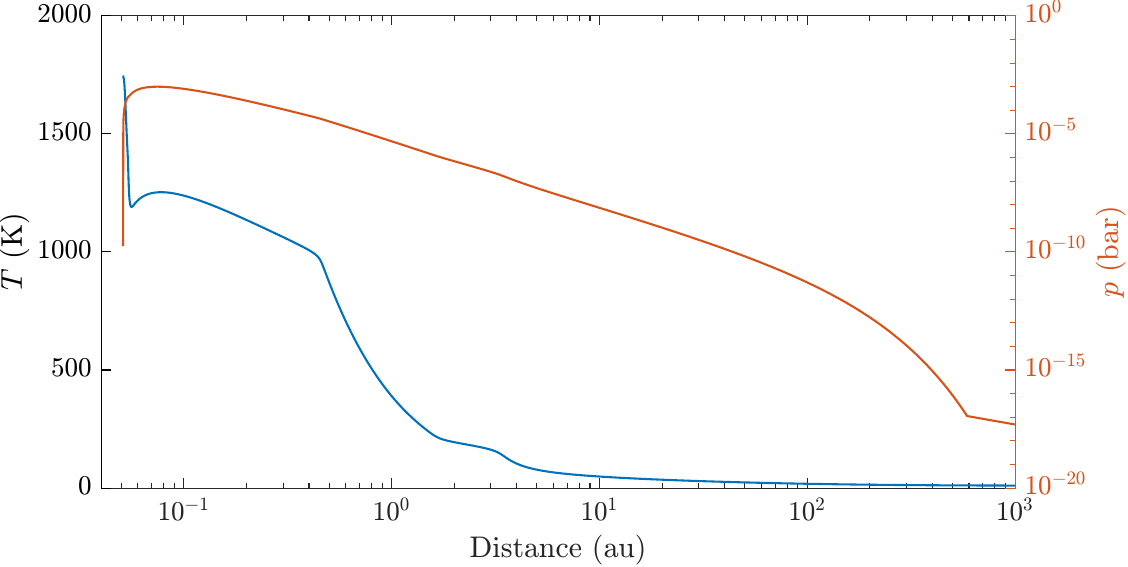}}
	\caption{Temperature (blue) and pressure (red) profiles of the midplane of a protoplanetary disk as a function of distance from the central star. The profiles are based on calculations with the disk model described in \citet{Emsenhuber2021A&A...656A..69E} and \citet{Weder2023A&A...674A.165W}.}
	\label{fig:disk_struct}
\end{figure}

In this section, we test our updated \fc code on a temperature-pressure structure of the midplane of a protoplanetary disk. The temperature and pressure profile is based on the disk model described in \citet{Emsenhuber2021A&A...656A..69E} and \citet{Weder2023A&A...674A.165W} and shown in Fig. \ref{fig:disk_struct}. 
Throughout the disk the pressure varies by more than 14 orders of magnitudes, from about $10^{-3}$ bar to $10^{-17}$ bar. The temperature starts at roughly 1700 K near the central star and drops to values of almost 10 K in the outermost part of the disk.
The low temperatures and densities, especially at larger distances, make this scenario numerically very challenging. In the outer parts of the disk, essentially all condensates considered in \fcc have initially activities larger than unity, which makes it obviously quite difficult to obtain the final set of stable condensates according to the phase rule. 

We note that this calculation is only done to test the numerical stability of the chemistry and condensation scheme and not to properly model the chemical composition of a protoplanetary disk. In reality the disk chemistry is rather complex, involving various additional chemical and physical processes, such as photoevaporation for example \citep[see, e.g.][]{2013ChRv..113.9016H}. For the calculations here, we also removed germanium from the list of considered elements, because currently Ge has no associated gas phase molecules or condensates due to the lack of corresponding data in the JANAF tables \citep{Chase1998}.

\begin{figure}
    \resizebox{\columnwidth}{!}{\includegraphics{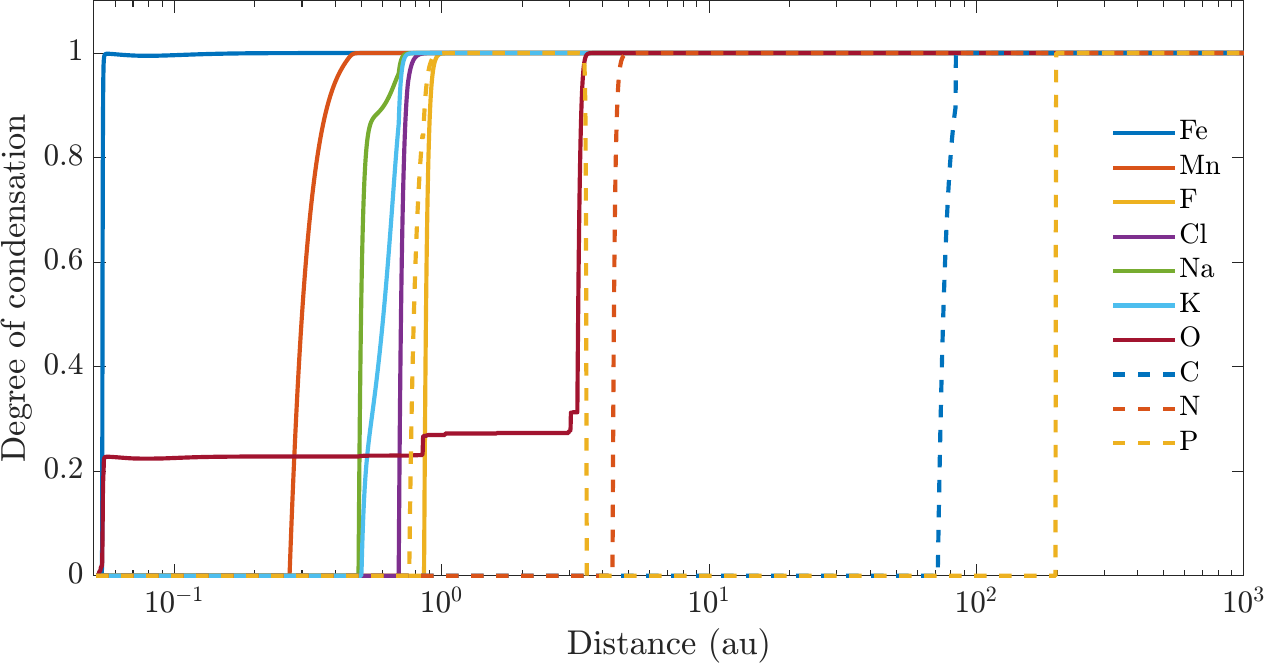}}\\
    \resizebox{\columnwidth}{!}{\includegraphics{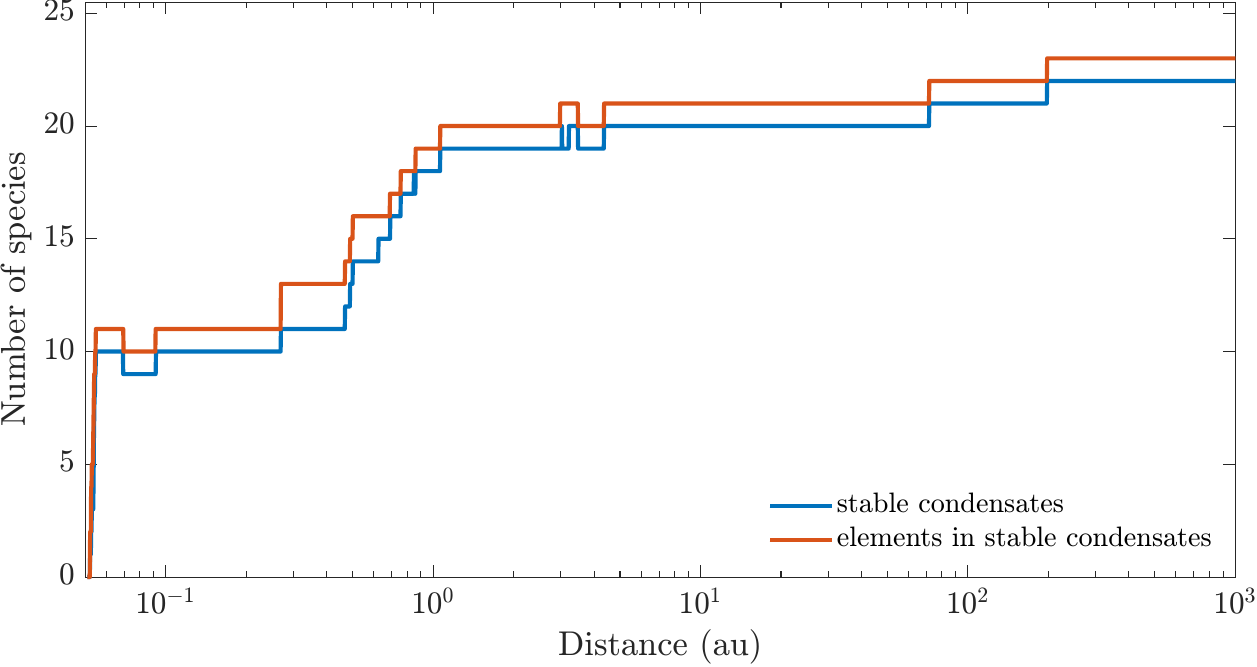}}
    \caption{Degree of condensation of selected elements (upper panel) and the number of stable condensates (lower panel) as a function of the distance from the central star for the protoplanetary disk. The lower panel also additionally shows the total number of elements contained in stable condensates to verify that the results satisfy the phase rule.}
    \label{fig:disk_nb_cond}
\end{figure}

In total, this calculation included 26 elements. Thus, following the phase rule a maximum of 25 different condensates should be able to co-exist at most. As the lower panel of Fig. \ref{fig:disk_nb_cond}  shows, the highest number of stable condensates in our calculation reaches a value of 22 in the outer disk (see Table \ref{tab:condensates_disk}), clearly satisfying the phase rule. The number of stable condensates should also not be higher than the number of elements contained in them. In our case, the converged system also satisfies this requirement. 

The upper panel of Fig.~\ref{fig:disk_nb_cond} shows the degree of condensation of selected elements as a function of distance (upper panel). Just like in the previous case of \citet{Sharp1990ApJS...72..417S}, elements tend to completely condense out below a certain temperature. Iron is lost from the gas phase very close to the inner disk boundary. The same also applies to other elements contained in high-temperature condensates, such as Ca or Al (not shown). More volatile elements, such as Mn, Na, K, or Cl start to condense at distances between 0.3 and 1 au. Oxygen is also contained in many of the high-temperatures condensates. Due to the high element abundance of O compared to the other elements only about 20\% of the element is lost from the gas phase at small distances, though. However, once the temperature-pressure profile crosses the water ice line of the protoplanetary disk, oxygen condenses out completely into \ch{H2O(s)} at distances larger than roughly 3.5 au. 

In the outer part of the disk, only hydrogen and the noble gases argon, neon, and helium remain in the gas phase. Hydrogen is also contained in some condensates, most notably water, ammonia, and methane ice. However, due to its very high element abundance, its degree of condensation never exceeds 0.2\%.

Phosphorus is the only element that has a distinctly non-monotonic degree of condensation. With increasing distance it condenses out close to 1 au, returns to the gas phase near 3 au and finally condenses out again at $r > 200$ au.

\begin{figure}
	\resizebox{\columnwidth}{!}{\includegraphics{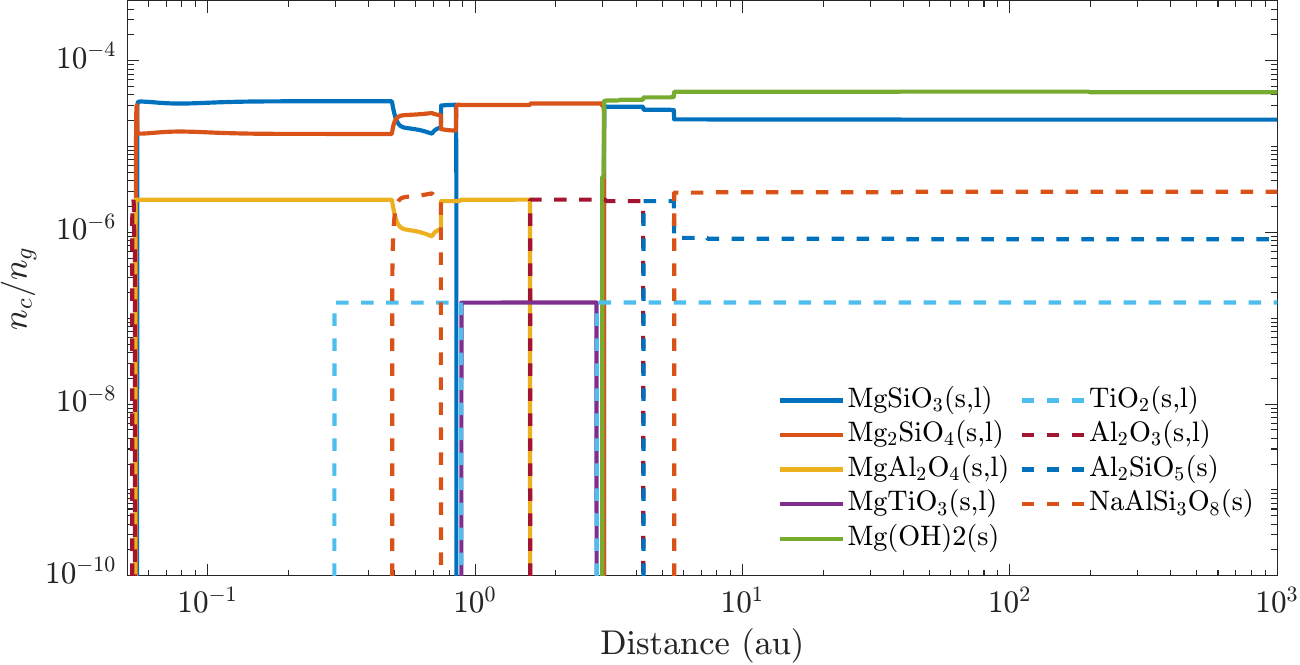}}\\
	\resizebox{\columnwidth}{!}{\includegraphics{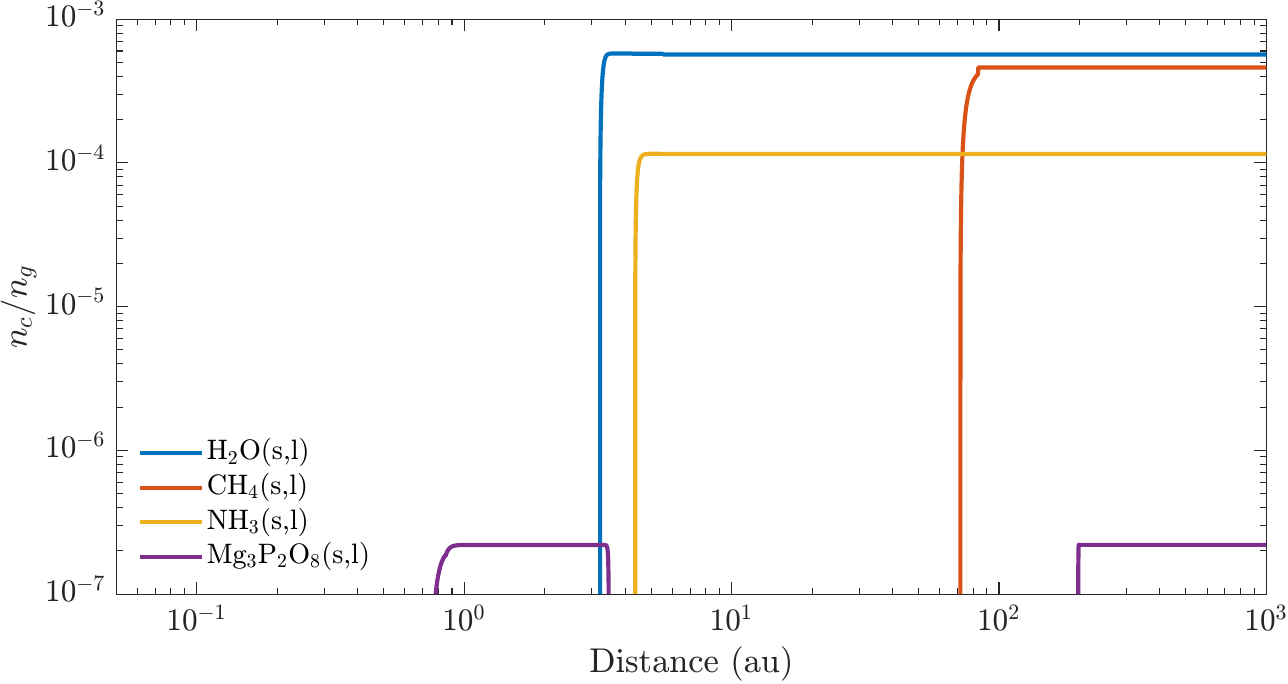}}	
	\caption{Abundance profiles of selected condensates as a function of distance in the protoplanetary disk model. The upper panel focuses on Mg, Ti, and Al-bearing refractory condensates, while the lower panel shows the more volatile condensate species.}
	\label{fig:disk_cond}
\end{figure}

Even though many refractory elements are almost completely bound in condensates from the inner to the outer regions of the disk, the actual condensates formed by them change with distance. Since many of the elements are found in multiple different condensates, they all compete with each other for the available elements. This is illustrated for a sample of magnesium, aluminum, and titanium-bearing condensates in the upper panel of Fig. \ref{fig:disk_cond}. 

With increasing distance from the disk's centre magnesium first condenses out into enstatite, forsterite, and spinel. Up to distances of 3 au, enstatite and forsterite compete for being the highest-abundance magnesium-bearing condensate. At 3 au, \ch{Mg(OH)2} forms and takes up most of the available magnesium. Consequently, forsterite is no longer able to exist, while enstatite is still able to co-exist as it requires smaller amounts of magnesium to form. 

Aluminum is first found predominantly in the form of spinel (\ch{MgAl2O4}). Near 0.5 au, it competes with \ch{NaAlSi3O8} for the element silicon and finally disappears close to 2 au in favour of corundum (\ch{Al2O3}). The latter, though, is only stable for a short range of distances and is then replaced by \ch{Al2SiO5}. This species can easily form there because the disappearance of forsterite in favour of \ch{Mg(OH)2} provides an abundant amount of silicon. As more silicon becomes available, the abundance of corundum is decreased again by the formation of \ch{NaAlSi3O8}, but still remains stable.

Titanium is predominantly found in only two condensates: \ch{TiO2} and \ch{MgTiO3}. Unlike magnesium or aluminium it does apparently not form several different condensates simultaneously. The innermost stable Ti-compound is \ch{TiO2}. It is replaced by \ch{MgTiO3} near 1 au. However, with the formation of \ch{Mg(OH)2} not enough magnesium is left and titanium returns to form \ch{TiO2}.

As shown in the bottom panel of Fig. \ref{fig:disk_cond}, at lower temperatures the volatile species \ch{H2O}, \ch{CH4}, and \ch{NH3} condense beyond 3 au, 4 au, and 70 au, respectively. These condensing species effectively remove the elements O, C, and N from the gas phase. The last element to leave the gas phase is phosphorus. As already noted above, P has a non-monotonic degree of condensation. Near 1 au, it first condenses out into \ch{Mg3P2O8}. With the formation of \ch{Mg(OH)2}, though, magnesium is no longer available. Since also no other phosphorus-bearing condensate has a high enough activity, phosphorus returns back to the gas phase. Only at much lower temperatures in the outer part of the disk, some magnesium becomes available again to form \ch{Mg3P2O8}, removing P from the gas phase once more. 

\begin{table}
  \caption[]{List of stable condensates in the outer parts of the protoplanetary disk with the associated elements according to the phase rule.}
  \label{tab:condensates_disk}
  \centering
  \begin{tabular}{lcl}
  \hline\hline
  Stable condensate & Associated element & $n_c/n_{c,\mathrm{max}}$\\
  \hline
    \ch{Al2SiO5(s)} & Al & 0.35\\ 
    \ch{NH4Cl(s)} & Cl & 1.0\\ 
    \ch{Co(s)} & Co & 1.0\\
    \ch{Cr2O3(s,l)} & Cr & 1.0\\
    \ch{Cu(s,l)} & Cu & 1.0\\
    \ch{MgF2(s,l)} & F & 1.0\\
    \ch{FeS(s,l)} & S & 0.9\\
    \ch{Mg(OH)2(s)} & Mg & 0.63\\
    \ch{MgSiO3(s,l)} & Si & 0.37\\
    \ch{Mg3P2O8(s,l)} & P & 1.0\\
    \ch{Ni3S2(s,l)} & Ni & 1.0\\
    \ch{TiO2(s,l)} & Ti & 1.0\\
    \ch{V2O3(s,l)} & V & 1.0\\
    \ch{Zn(s,l)} & Zn & 1.0\\
    \ch{MnS(s)} & Mn & 1.0\\
    \ch{Fe2SiO4(s)} & Fe & 0.63\\
    \ch{CaMgSi2O6(s)} & Ca & 1.0\\
    \ch{KAlSi3O8(s)} & K & 1.0\\
    \ch{NaAlSi3O8(s)} & Na & 1.0\\
    \ch{H2O(s,l)} & O & 0.68\\
    \ch{CH4(s,l)} & C & 1.0\\
    \ch{NH3(s,l)} & N & 0.99\\
  \hline
  \end{tabular}
\end{table}

As discussed in Sect. \ref{sec:gibbs_phase_rule}, the phase rule and the requirement that the resulting system of activity equations for the set of stable condensates needs to be linearly independent limits the number of stable condensates. The latter requirement, however, also suggests that each condensed species should be linked with a related element that is associated with its stability. Often, this should be the least abundant element contained in a condensate normalised with the corresponding stoichiometric coefficient (see also Appending \ref{sec:puzzle}).

To illustrate this, we focus on the condensates present at the outermost grid point of the protoplanetary disk. According to Fig. \ref{fig:disk_nb_cond}, this point has the highest possible number of elements contained in condensates. All other elements left in the gas phase are the noble gases (He, Ne, and Ar) that don't have any condensate species associated with them in \fcc. In Table \ref{tab:condensates_disk} we list all 22 condensate species that are stable in the outer part of the disk. We also list the associated element for each condensate and the fraction of the fictitious condensate density $n_c$ and its largest possible value $n_{c,\mathrm{max}}$ (see Eq. \eqref{eq:cond_max_density}). If $n_c$ is smaller than $n_{c,\mathrm{max}}$, the linked element is contained in more than one condensate. The assignment of the condensates to their respective elements is done by the procedure described in Appendix \ref{sec:puzzle}. 

The table clearly suggests that each element is uniquely linked with one specific condensate. Even though some elements, such as Mg or Al, can form several different condensates simultaneously they are the associated species in only a single condensate. In many cases, the associated element is also the least abundant one in a specific condensate, often containing the total available inventory of that element. This also explains why, for example, the titanium-bearing condensates form very distinct condensation sequences (see Figs. \ref{fig:sh_cond_seq} \& \ref{fig:disk_cond}). Species like \ch{TiO2(s,l)}, \ch{Ti2O5(s,l)}, or \ch{MgTiO3(s,l)} cannot exist simultaneously as Ti would be the associated and least abundant element in all of them. Magnesium, on the other hand, can also be present in other condensate species besides \ch{Mg(OH)2(s)}, such as \ch{MgSiO3(s,l)} in this example here, because it cannot fully condense out in a single condensate due to its higher abundance compared to the other elements in these condensates. Thus, elements with higher element abundances will usually tend to form multiple different condensates if those also contain elements with smaller abundances.

\subsection{Equilibrium condensation vs. rainout approximation}
\label{sec:rainout_example}

Finally, we apply the \fc code to an atmosphere of a typical brown dwarf. As discussed in Sect. \ref{sec:rainout}, for atmospheres of planets and brown dwarfs the rainout approximation is a more suitable approach to calculate the chemical composition than the pure equilibrium condensation used in the previous two examples. In the following, we will show the impact of these two different approaches on the forming condensates and the chemical species remaining in the gas phase

\begin{figure}
    \centering
	\resizebox{0.7\columnwidth}{!}{\includegraphics{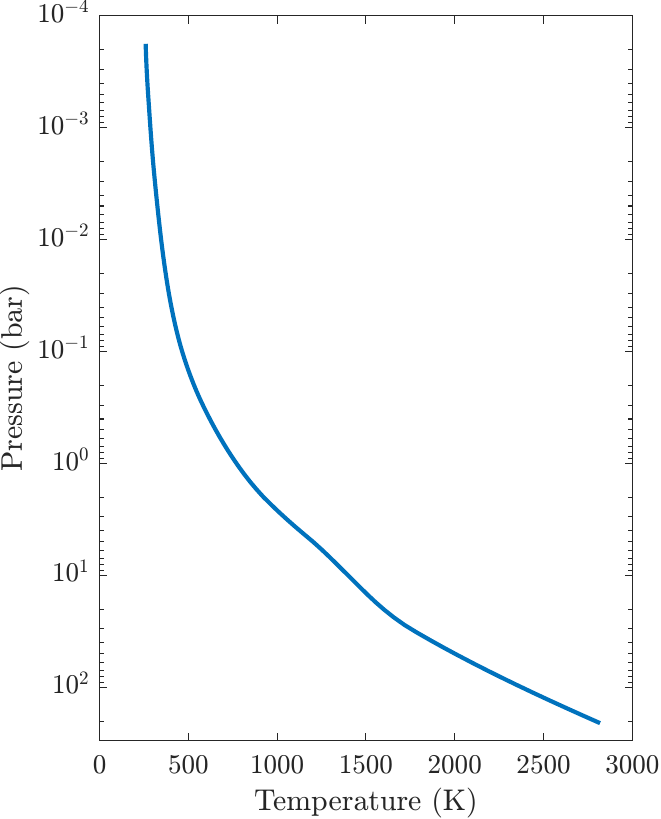}}
	\caption{Temperature-pressure profile of a typical T5 brown dwarf used for the chemistry calculations. The profile is taken from the grid of \citet{Marley2021ApJ...920...85M} for an effective temperature of 750 K and a surface gravity of $g = 3.16\times10^{4}\,\mathrm{cm}\,\mathrm{s}^{-2}$ ($\log g = 4.5$) .}
	\label{fig:bd_temperature}
\end{figure}

For the atmospheric temperature-pressure structure of the brown dwarf we use output from the \texttt{Sonora} grid of brown dwarf models published by \citet{Marley2021ApJ...920...85M}, available on Zenodo\footnote{\url{https://zenodo.org/record/5063476}}. In particular, we choose a model with an effective temperature of 750 K and a surface gravity of $g = 3.16\times10^{4}\,\mathrm{cm}\,\mathrm{s}^{-2}$ ($\log g = 4.5$), resembling a typical T5 dwarf. The corresponding temperature-pressure profile is shown in Fig. \ref{fig:bd_temperature}. Based on the general understanding of brown dwarf atmospheres, we expect the lower atmosphere to be dominated by the condensation of iron and various silicates, while in the upper atmosphere species like \ch{Na2S}, \ch{NaCl}, or \ch{KCl} should appear (see \citet{Marley2013cctp.book..367M} or \citet{Morley2012ApJ...756..172M}, for example).

The temperature-pressure profile is used as an input for two different chemistry calculations with \fcc, the first one using the standard equilibrium condensation approach from the two previous test cases and a second case with the rainout approximation. As described in Sect. \ref{sec:rainout}, for the latter one we start the chemistry calculations at the bottom of the atmosphere and proceed upwards towards lower pressure, while calculating the gas phase composition and the condensed phase, respectively. Once condensation is encountered, we use the resulting fictitious number densities $n_c$ of the condensates to calculate the effective abundances $\phi_j$ of these elements left in the gas phase. These new, reduced element abundances are then used in the subsequent calculations of the upper atmosphere.

\begin{figure*}
    \centering
	\resizebox{0.65\columnwidth}{!}{\includegraphics{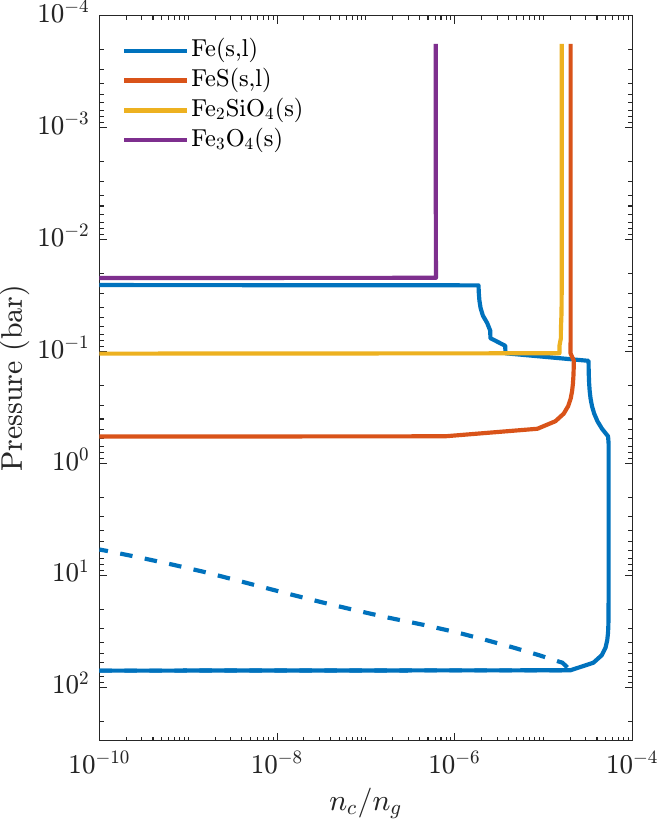}}
	\resizebox{0.65\columnwidth}{!}{\includegraphics{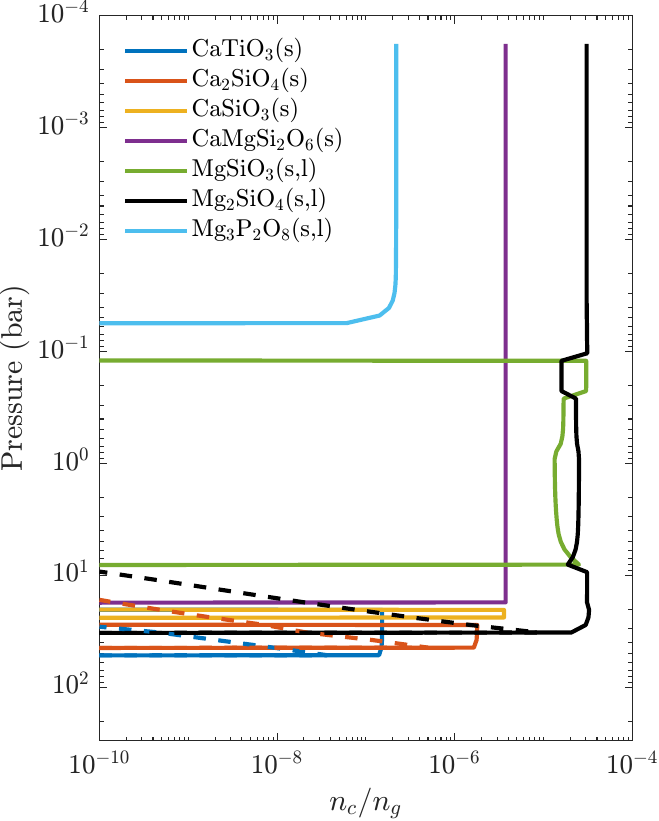}}
        \resizebox{0.65\columnwidth}{!}{\includegraphics{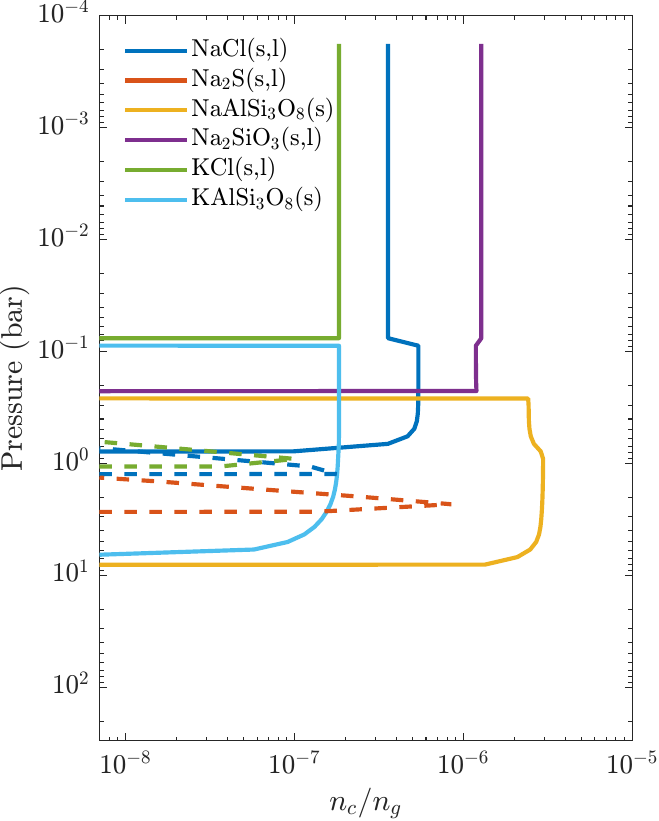}}\\
        \resizebox{0.65\columnwidth}{!}{\includegraphics{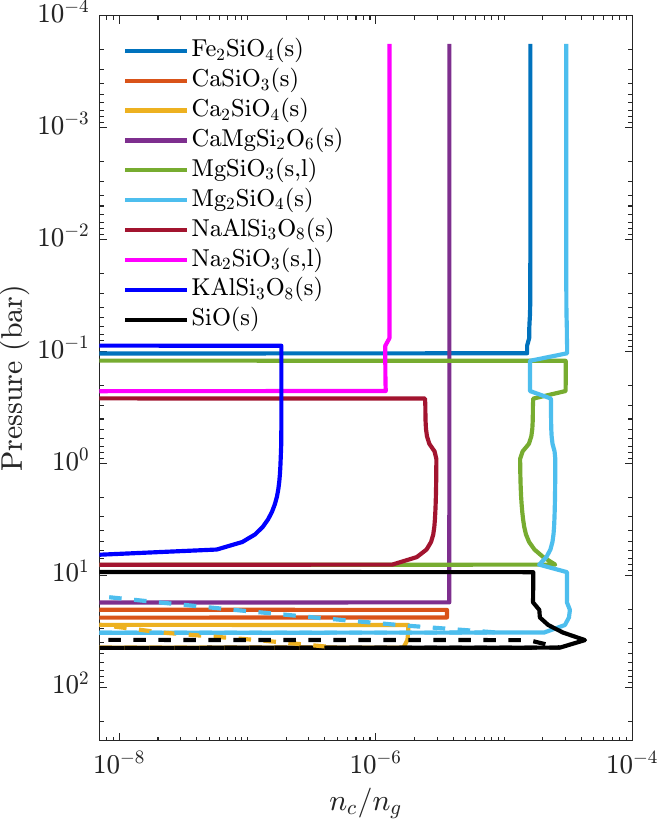}}
        \resizebox{0.65\columnwidth}{!}{\includegraphics{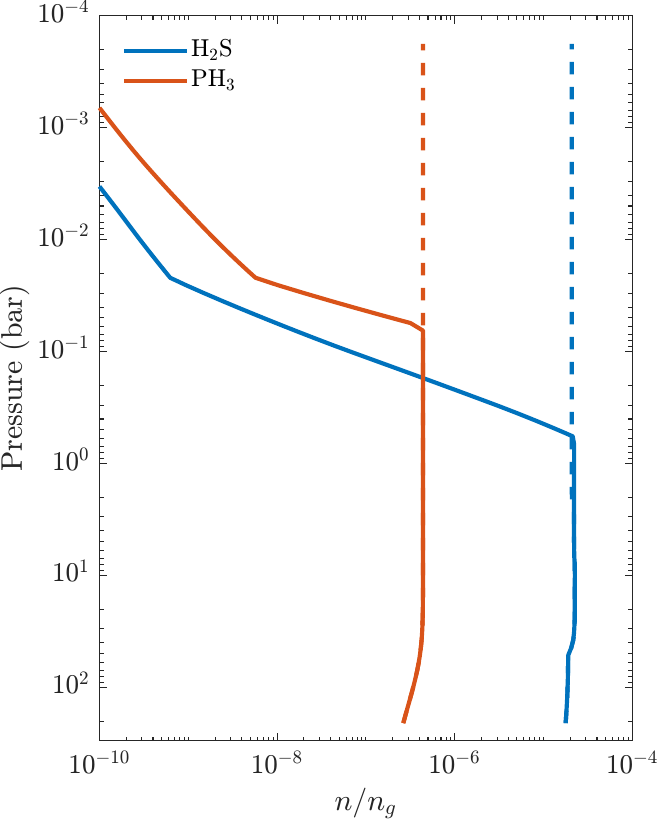}}
        \resizebox{0.63\columnwidth}{!}{\includegraphics{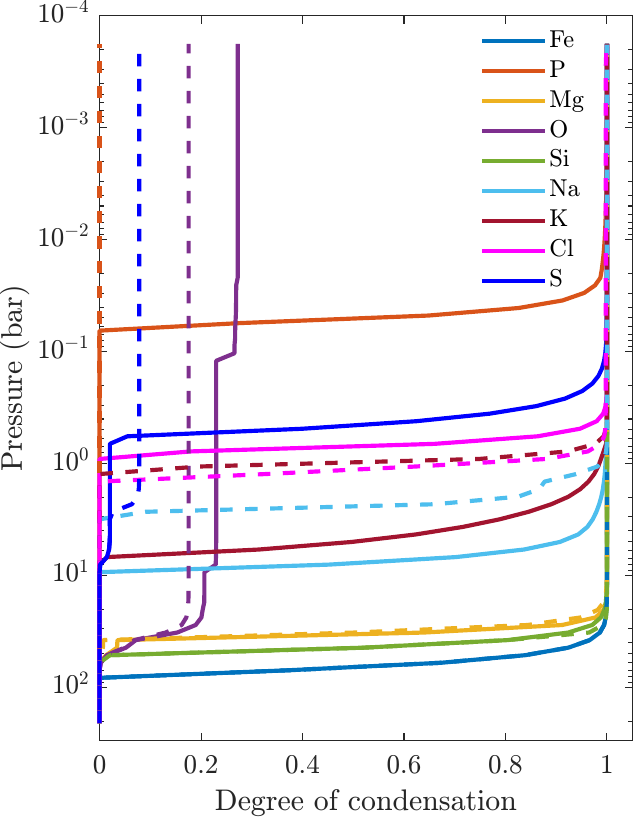}}
	\caption{Impact of condensation on the chemistry of a brown dwarf atmosphere. The top panel shows abundance profiles for selected condensates: iron-bearing condensates (left), magnesium and calcium species (middle), sodium and potassium condensates (right). Bottom panel: abundance profiles for silicon-bearing condensates (left), impact of the two different condensation approaches on the two gas-phase species hydrogen sulfide (\ch{H2S}) and phosphine (\ch{PH3}) (middle), the degree of condensation for selected elements (right). The solid lines refer to the equilibrium condensation calculations. Dashed lines indicate the results for the rainout approximation.}
	\label{fig:bd_cond}
\end{figure*}

In the top panel of Fig. \ref{fig:bd_cond} we show some important iron, calcium, and magnesium-bearing condensates comparing both condensation descriptions. Iron is one of the first condensates to become stable at a pressure of roughly 70 bar and a temperature of about 2200 K. Above this pressure no condensate currently included in \fcc can stably exist because of the high temperatures at the bottom of the atmosphere. For the equilibrium condensation, iron is found in the form of \ch{Fe(s,l)} in the lower atmosphere, while in the upper atmosphere the dominant iron-bearing condensates are fayalite (\ch{Fe2SiO4(s)}), magnetite (\ch{Fe3O4(s)}), and \ch{FeS(s)}. 

In contrast to that, the condensation into \ch{Fe(s,l)} at pressures of about 10 bar already depletes the gas phase iron element abundance in the upper atmosphere for the rainout approximation. Consequently, no other Fe compounds can form there as essentially no iron is left in the gas phase at pressures lower than about 4 bar. The rainout of iron into an \ch{Fe(s,l)} condensate layer and, consequently, its absence in the upper atmosphere agrees with results previously described in studies such as \citet{Marley2021ApJ...920...85M} or \citet{Visscher2010ApJ...716.1060V}.

The corresponding results for major calcium and magnesium-bearing condensates are depicted in the top middle panel of Fig. \ref{fig:bd_cond}. Here we can clearly see that in the equilibrium condensation case, calcium is first condensed in \ch{CaTiO3(s)} as well as \ch{Ca2SiO4(s)}. Around 20 bar Ca is briefly mostly contained in \ch{CaSiO3(s)}, while the upper atmosphere is then dominated by \ch{CaMgSi2O6(s)}. 

The latter two condensates do not form in the rainout approach, however. Here, Ca is depleted in the lower atmosphere by condensation into \ch{CaTiO3(s)} and \ch{Ca2SiO4(s)}. As a result, more complex, secondary condensates like \ch{CaMgSi2O6(s)} cannot form in the upper atmosphere.

Magnesium is mostly contained in the two condensates forsterite (\ch{Mg2SiO4(s,l)}) and enstatite (\ch{MgSiO3(s,l)}). Condensation of forsterite takes place at pressures below the 20 bar level in the equilibrium condensation calculation, followed by \ch{MgSiO3(s,l)} between 10 bar and 0.1 bar. Both condensates co-exist in this pressure range, competing both for the available magnesium. Below pressures of 0.1 bar, however, \ch{MgSiO3(s,l)} stops forming, while forsterite becomes the dominating magnesium-bearing condensate. The second-most abundant magnesium condensate assuming the equilibrium condensation approach in the upper atmosphere is \ch{Mg3P2O8(s,l)}.

In the rainout approach, on the other hand, enstatite does not form. Condensation of magnesium into forsterite near 20 bar already depletes the gas phase of magnesium in the upper atmosphere, such that \ch{MgSiO3(s,l)}, as well as other Mg-bearing species are unable to form.

The top right panel of Fig. \ref{fig:bd_cond} shows the condensation of sodium and potassium-bearing species. In the equilibrium condensation calculation, Na first condenses into albite (\ch{NaAlSi3O8(s)}) near a pressure of 9 bar and into halite (\ch{NaCl(s)}) below 1 bar. Around 0.2 bar, albite is replaced by sodium silicate (\ch{Na2SiO3(s,l)}) as the most abundant sodium-bearing condensate. Potassium is only found in two stable condensates: microcline (\ch{KAlSi3O8(s)}) in the lower atmosphere and \ch{KCl(s)} (sylvite) in the upper part.

In the rainout approach model, albite is unable to form because neither silicon nor aluminum are available as they have already been depleted in the deeper atmosphere. Consequently, sodium stays in the gas phase at lower pressures compared to the equilibrium condensation case and finally condenses out into \ch{Na2S} at about 3 bar and \ch{NaCl} near 1 bar. At roughly the same pressure level, potassium rains out into a sylvite layer without forming an additional condensate species.

The condensation sequence of silicon-bearing species is depicted in the bottom left panel of Fig. \ref{fig:bd_cond}. As the figure suggests, this sequence is rather complicated, with silicon being present in as much as five different condensates simultaneously. In the lower atmosphere, Si is mostly contained in Ca-bearing condensates and \ch{SiO(s)} whereas in the upper atmosphere it is predominantly chemically bound in forsterite, fayalite, diopside, and \ch{Na2SiO3(s,l)}. The absence of quartz (\ch{SiO2(s)}) as a stable condensate in the equilibrium condensation approach is consistent with studies such as \citet{Visscher2010ApJ...716.1060V}, though it has been predicted to form under non-equilibrium conditions \citep{Helling2006A&A...455..325H}. In the rainout approximation, silicon rains out already in the deeper atmosphere into three distinct condensates: \ch{SiO(s)}, larnite (\ch{Ca2SiO4(s)}), and forsterite.

The two different condensation treatments do not only have an impact on the condensing species, but also on the chemical composition of the gas phase. Examples for two gas-phase molecules are shown in the bottom middle panel of Fig. \ref{fig:bd_cond}. The most well-known example is the impact on \ch{H2S}, as already extensively discussed in the literature \citep[see][for example]{Lodders2002Icar..155..393L}. As already discussed, in the equilibrium condensation approach, iron is mostly contained in \ch{FeS(s,l)} in the upper atmosphere, depleting the gas phase of both iron and sulphur. As a result, the \ch{H2S} molecule is unable to efficiently form in the gas phase in the upper atmosphere. 

In contrast to the equilibrium condensation approach, we can see that \ch{H2S} is abundantly present in case of the rainout approximation. This is caused by the aforementioned depletion of iron into \ch{Fe(s,l)} in the lower atmosphere. As a result, \ch{FeS(s,l)} is unable to form in the rainout case and sulphur remains available to form \ch{H2S}.

Another gas-phase species strongly affected by the differences of the two condensation treatments is phosphine (\ch{PH3}) as suggested by the results shown in Fig. \ref{fig:bd_cond}. In the equilibrium condensation approach, phosphorus condenses into \ch{Mg3P2O8(s,l)} in the upper atmosphere, as depicted in Fig. \ref{fig:bd_cond}. However, this condensate cannot be formed in the calculation with the rainout approximation because Mg is removed from the gas phase in the deeper atmosphere. Consequently, enough phosphorus remains in the gas phase to form phosphine.

The impact of the two different condensation descriptions can also seen in the form of the element depletion, depicted in the bottom right panel of Fig. \ref{fig:bd_cond}. While the degree of condensation for elements in high-temperature condensates, such as Fe, Si, or Mg, is not strongly affected by the condensation treatment, larger differences can be noted in the aforementioned elements S and P. For the equilibrium condensation approach, both condense out at 1 bar and about 0.1 bar, respectively. In the rainout scenario, however, only about 10\% of sulphur actually condenses into \ch{Na2S(s,l)}, while the degree of condensation for phosphorous remains zero throughout the entire atmosphere. Other elements, such as sodium or potassium, condense out completely in both condensation approaches, albeit they leave the gas phase at considerable lower pressures for the rainout approximation. Conversely, chlorine already rains out at higher pressures than in the equilibrium condensation calculation.

\section{Summary}

In this study we introduce a new version of \fc. With a version number of 3.0, we refer to this updated code as \fcc. While the two previous versions of \fc only accounted for the gas phase composition, the new version now introduces the treatment of condensates to the code. In particular, \fcc can compute scenarios involving equilibrium condensation as well as the rainout approximation often used in the modelling of atmospheres of (exo)planets and brown dwarfs. Together with the code update we also add about 290 liquid and solid condensate species to \fc.

The numerical treatment of condensates is based on the basic ideas presented by \citet{Leal2016AdWR...96..405L} and have been successfully adapted to the numerical formalism used in \fc. A major advantage compared to other established equilibrium condensation codes in the field of astrophysics is the automatic selection of stable condensates satisfying the phase rule.

The updated version of \fc is tested for several scenarios involving condensation. In particular, we reproduce the equilibrium condensation calculations by \citet{Sharp1990ApJS...72..417S} and also test \fcc for the numerically very challenging scenario of the midplane of a protoplanetary disk. Finally, we use a theoretical atmosphere model of a typical T5 brown dwarf to showcase the differences between equilibrium condensation and the rainout approximation.

\fcc is programmed in object-oriented C++ and offers the additional Python module \textsc{pyFastChem} that allows it to be run directly from any Python script. The code is released as open source at GitHub (\url{https://github.com/exoclime/FastChem}) under the GNU General Public License version 3 \citep{Gnu07}. The repository also contains several Python example scripts that showcase the use of \fcc under Python, as well as an extensive manual that provides detailed descriptions of the code, its required inputs and available, optional parameters.

\section*{Acknowledgements}

The authors would like to dedicate this publication to the memory of Erwin Sedlmayr who passed away in January 2022.
D.K. acknowledges financial support from the Center for Space and Habitability (CSH) of the University of Bern. This work has been carried out within the framework of the NCCR PlanetS supported by the Swiss National Science Foundation under grants 51NF40\_182901 and 51NF40\_205606.
D.K. would like to thank Jesse Weder and Christoph Mordasini for providing the temperature-pressure structure of the protoplanetary disk used in this study.

\section*{Data Availability}

The data underlying of this paper are available in the \fc GitHub repository, at \url{https://github.com/exoclime/FastChem}.

\bibliographystyle{mnras}
\bibliography{references}

\appendix

\section{Data fits}
\label{sec:data_fits}

The calculation of the condensed phase requires some important thermochemical data as basic input, namely the temperature-depend equilibrium constants. The dimension-less equilibrium constant $\bar K_c(T)$ for a specific condensate $c$ is calculated from the Gibbs free energy of reaction \citep[e.g.,][]{Stock2018MNRAS.479..865S}
\begin{equation}
   \ln \bar K_c(T) = -\frac{\Delta_\mathrm{r} G_c^\standardstate(T)}{R T} \ ,
   \label{eq:lnk_fastchem}
\end{equation}
where $\Delta_\mathrm{r} G_c^\standardstate$ is the change of Gibbs free energy for the dissociation reaction 
\begin{equation}
  \ch{A_aB_bC_c(s) <=> a A(g) + b B(g) + c C(g)} \ ,
\end{equation}
of a condensate species into its elements, given by:
\begin{equation}
  \begin{split}
   \Delta_\mathrm{r} G_c^\standardstate(T) &= \Delta_\mathrm{f} G_c^\standardstate(T) -  \sum_{j \in \mathcal E} \nu_{cj} \Delta_\mathrm{f} G_{j}^\standardstate(T)  \\
                                           &= G_c^\standardstate(T) -  \sum_{j \in \mathcal E} \nu_{cj} G_{j}^\standardstate(T) \ , \qquad\qquad c \in \mathcal C \ .
  \end{split}
  \label{eq:delta_gr}
\end{equation}
Here, the $\Delta_\mathrm{f} G_c^\standardstate(T)$ are the Gibbs free energies of formation and the $G_c^\standardstate(T)$ the corresponding Gibbs free energies.
It is important to note that the tabulated thermochemical data, such as the $\Delta_\mathrm{f} G^\standardstate(T)$ or the $\ln \bar K$, in the corresponding databases are always given with respect to the elements in a specific reference state. For example, the JANAF tables \citep{Chase1998} use in general the chemically most stable form of an element at standard pressure and a temperature of 298.15 K as the reference state. Thus, the thermochemical data of, for example, \ch{CO2} is given with respect to \ch{C(s)} and \ch{O2} in the JANAF tables because the stable form of oxygen at 1 bar and 298.15 K is molecular oxygen while carbon's is graphite. By calculating the equilibrium constants via Eqs. \eqref{eq:lnk_fastchem} \& \eqref{eq:delta_gr} with \fc, we effectively adjust the reference state of the elements from the tabulated data to their monatomic form. Thus, the $\ln \bar K$ used in \fc and given in Table \ref{tab:condensate_fits} should not be confused with the corresponding $\ln \bar K$ tabulated in thermochemical databases, such as the JANAF tables. Data from different sources, therefore, generally cannot be used interchangeably with \fc and usually need to be converted to the reference state used in our formalism.

If the number densities $n_j$ are given in units of cm$^{-3}$, then in the activity equation \eqref{eq:activity_eq} for the condensates $K_c$ has units of 
\begin{equation}
  \left[K_c \right] = \left( \mathrm{cm}^{-3} \right)^{- \sum_{j \in \mathcal E} \nu_{cj}} \ .
\end{equation}
It is related to the dimensionless mass action constant $\bar K_c(T)$ by
\begin{equation}
    \ln K_c(T) = \ln \bar K_c(T) - \ln \left( \frac{p^\standardstate}{k_{\mathrm B} T} \sum_{j \in \mathcal E} \nu_{cj} \right)  \ ,
\end{equation}
where $p^\standardstate$ is the pressure of the standard state of the thermochemical data. We note that this relation explicitly makes use of the ideal gas law. For the data used in the \fc code, $p^\standardstate$ is equal to 1 bar.

As stated by \citet{Stock2018MNRAS.479..865S}, we parameterise $\ln \bar K_c$ as a function of the temperature $T$ in form of the polynomial
\begin{equation}
    \ln \bar K_c(T) = \frac{a_0}{T} + a_1 \ln T + b_0 + b_1 T + b_2 T^2 \ ,
    \label{eq:lnk_fit}
\end{equation}
where $a_i$ and $b_i$ are the corresponding fit coefficients that are tabulated in the \fc input file for each species.

Following the approach of the gas-phase data in the previous versions of \fc, we base our choice of species mainly on the JANAF tables \citep{Chase1998}. In particular, we use all condensates for the elements contained in \fc included in the JANAF tables. Species that have both liquid and solid-form data are fitted independently. Thus, some condensate species have two different sets of polynomial coefficients attached to them. \fcc in principle allows a condensate with a certain stoichiometric formula to have even more sets of fit coefficients. It is, thus, possible to fit also different crystal phases independently if so desired by the user. A list of all species and the fit coefficients can be found in Table \ref{tab:condensate_fits}.

In addition to the JANAF tables, we also add selected condensates from other data sources. This includes condensate species from \citet{Sharp1990ApJS...72..417S} not available in the JANAF tables, some volatile, low-temperature condensates such as water, ammonia, methane, or carbon dioxide, as well as silicon monoxide (\ch{SiO(s)}) from \citet{Gail2013AA...555A.119G}. 

For species where the thermochemical data is given in terms of the saturation vapour pressure $p_{\mathrm{vap},c}(T)$ rather than the Gibbs free energy, we use the relation
\begin{equation}
    \ln \left( \frac{p_{\mathrm{vap},c}(T)}{p^\standardstate} \right) =  \frac{G_c^\standardstate(T) - G_i^\standardstate(T)}{RT}
\end{equation}
to calculate $G_c^\standardstate(T)$, where $G_i^\standardstate(T)$ is the Gibbs free energy of the corresponding molecule. For the latter we again use the gas-phase data from the JANAF tables.

We note that we exchanged the data for \ch{SiO2(s,l)} from the JANAF tables with that from \citet{Barin1995}. The JANAF data suggests that \ch{SiO2(l)} would switch into solid \ch{SiO2}, more specifically quartz, at a temperature of 1696 K. However, the standard phase diagram of \ch{SiO2} as well as the data from \citet{Barin1995} or the NASA Glenn coefficients \citep{McBride2002} all agree that at standard pressure, \ch{SiO2(l)} first transitions into the solid cristobalite at a temperature of 1996 K before its stable form finally changes to quartz at lower temperatures. Due to the discrepancies between the JANAF tables and the other literature, we choose to use the data from \citet{Barin1995} instead.

Thermochemical data for condensates are often only available over a very restricted temperature range. Therefore, by default \fcc avoids extrapolating the fit equation \eqref{eq:lnk_fit} outside of the temperature range of the underlying data. A special configuration parameter is available in \fcc, though, that allows the user to override this limitation. However, this parameter should only be used with great care as the extrapolation might result in an unphysical, non-monotonic behaviour of the $\ln K_c$ fit for some condensate species. 

\begin{table*}
  \caption[]{An overview of condensates and the fit coefficients for the equilibrium constants included in \fc. Species with both liquid and solid phases have two separate sets of coefficients. The first line always refers to the solid species, while the second line represents the liquid phase. It is important to note that the tabulated $\ln \bar K$ use the elements in their monatomic form as reference state (see text for details).}
  \label{tab:condensate_fits}
  \centering
  \scriptsize
  \begin{tabular}{llcccccc}
  \hline\hline
  Species              & Name                      & $a_0$       & $a_1$        & $b_0$        &  $b_1$      &  $b_2$       & References\\
  \hline
\ch{Al(s)} & Aluminum & 3.956147e+04 & -8.786728e-01 & -1.167603e+01 & 2.133162e-03 & -3.413406e-07 & 1\\ 
\ch{Al2O3(s,l)} & Corundum & 3.684822e+05 & -1.041942e+01 & -2.792435e+01 & 1.179042e-02 & -1.587869e-06 & 1\\ 
 & Aluminum Oxide & 3.713491e+05 & 5.273685e+00 & -1.353453e+02 & 1.681739e-03 & -8.802596e-08 & \\ 
\ch{Al2S3(s)} & Aluminum Sulfide & 2.563027e+05 & -7.095578e+00 & -4.462246e+01 & 1.119856e-02 & -1.975854e-06 & 1\\ 
\ch{Al2SiO5(s)} & Aluminum Silicate, Kyanite & 5.918298e+05 & -1.467960e+01 & -5.539231e+01 & 1.428258e-02 & -1.532276e-06 & 1\\ 
\ch{Al4C3(s)} & Aluminum Carbide & 4.407611e+05 & -1.036553e+01 & -6.167107e+01 & 1.118864e-02 & -1.248269e-06 & 1\\ 
\ch{Al6Si2O13(s)} & Aluminum Silicate, Mullite & 1.547250e+06 & -3.695282e+01 & -1.487433e+02 & 3.637935e-02 & -4.036965e-06 & 1\\ 
\ch{AlCl3(s,l)} & Aluminum Chloride & 1.675202e+05 & -7.550810e+00 & -2.652395e+01 & 2.274542e-02 & -1.018958e-05 & 1\\ 
 &  & 1.652902e+05 & 2.464703e+00 & -7.571900e+01 & 1.931285e-03 & -2.825885e-07 & \\ 
\ch{AlClO(s)} & Aluminum Chloride Oxide & 1.794890e+05 & -1.203782e+00 & -4.577195e+01 & 2.044840e-03 & -1.880469e-07 & 1\\ 
\ch{AlF3(s,l)} & Aluminum Fluoride & 2.487414e+05 & -5.745158e+00 & -3.427156e+01 & 7.085042e-03 & -9.144723e-07 & 1\\ 
 &  & 2.360574e+05 & -3.743293e+00 & -3.872272e+01 & 2.682689e-03 & -1.350757e-07 & \\ 
\ch{AlN(s)} & Aluminum Nitride & 1.339375e+05 & -3.463105e+00 & -1.434411e+01 & 3.453076e-03 & -3.929430e-07 & 1\\ 
\ch{C(s)} & Graphite & 8.568812e+04 & -1.859707e+00 & -6.365415e+00 & 1.052982e-03 & -6.538803e-08 & 1\\ 
\ch{CH4(s,l)} & Methane & 1.988879e+05 & -2.100289e+00 & -3.215642e+01 & -5.115678e-02 & 4.954560e-05 & 2\\ 
 &  & 1.993153e+05 & 1.153444e+01 & -8.945224e+01 & -1.628836e-01 & 2.023426e-04 & \\ 
\ch{CO(l)} & Carbon Monoxide & 1.298145e+05 & 2.919441e+00 & -3.273636e+01 & -1.572012e-02 & -5.510555e-05 & 3\\ 
\ch{CO2(s,l)} & Carbon Dioxide & 1.991499e+05 & 1.438951e+02 & -5.973862e+02 & -1.693532e+00 & 2.816767e-03 & 4\\ 
 &  & 2.008946e+05 & 6.846712e+01 & -3.867496e+02 & -2.638778e-01 & 1.596368e-04 & \\ 
\ch{Ca(OH)2(s)} & Calcium Hydroxide & 2.504859e+05 & -1.074866e+01 & -1.225092e+01 & 1.759083e-02 & -4.411867e-06 & 1\\ 
\ch{Ca(s,l)} & Calcium & 2.149529e+04 & 2.238090e-01 & -1.537980e+01 & 2.557372e-04 & 3.500411e-07 & 1\\ 
 &  & 2.038206e+04 & 6.678459e-01 & -1.739738e+01 & 6.213603e-04 & -6.114669e-08 & \\ 
\ch{Ca2Al2SiO7(s)} & Gehlenite & 8.690139e+05 & -8.773045e-01 & -2.108860e+02 & 3.890910e-03 & -1.420233e-07 & 5\\ 
\ch{Ca2SiO4(s)} & Larnite & 4.938758e+05 & -8.773129e-01 & -1.154846e+02 & 2.596067e-03 & -1.420238e-07 & 5\\ 
\ch{CaAl2Si2O8(s)} & Anorthite & 9.440506e+05 & -1.961755e+00 & -2.236517e+02 & 4.061832e-03 & -1.246940e-07 & 5\\ 
\ch{CaCl2(s,l)} & Calcium Chloride & 1.463160e+05 & -7.420422e-01 & -4.262592e+01 & 3.394191e-03 & -8.074484e-07 & 1\\ 
 &  & 1.451897e+05 & 2.554356e+00 & -6.282319e+01 & 1.077119e-03 & -8.936442e-08 & \\ 
\ch{CaF2(s,l)} & Calcium Fluoride & 1.875412e+05 & -2.241695e+00 & -3.559303e+01 & 3.779908e-03 & -2.709703e-07 & 1\\ 
 &  & 1.848023e+05 & 2.379252e+00 & -6.409940e+01 & 9.778805e-04 & -7.709693e-08 & \\ 
\ch{CaMgSi2O6(s)} & Diopside & 7.148791e+05 & -2.026835e+00 & -1.670276e+02 & 3.403319e-03 & -1.557866e-07 & 5\\ 
\ch{CaO(s,l)} & Lime & 1.274223e+05 & -1.487077e+00 & -2.452561e+01 & 2.263440e-03 & -2.336948e-07 & 1\\ 
 & Calcium Oxide & 1.335876e+05 & 1.154889e+01 & -1.210960e+02 & -1.932018e-03 & 4.857311e-08 & \\ 
\ch{CaS(s)} & Calcium Sulfide & 1.115334e+05 & -6.115886e-01 & -2.869597e+01 & 1.525313e-03 & -1.542806e-07 & 1\\ 
\ch{CaSiO3(s)} & Wollastonite & 3.627167e+05 & -9.463588e-01 & -8.280477e+01 & 1.787685e-03 & -8.890603e-08 & 5\\ 
\ch{CaTiO3(s)} & Perovskite & 3.674183e+05 & -1.766903e-01 & -8.730090e+01 & 1.522505e-03 & -6.177887e-08 & 5\\ 
\ch{ClSSCl(l)} & Sulfur Chloride & 1.039544e+05 & 3.853606e+00 & -7.902839e+01 & 2.217898e-05 & 9.895276e-08 & 1\\ 
\ch{Co(s,l)} & Cobalt & 5.131664e+04 & -1.624057e-01 & -1.718128e+01 & 4.248043e-04 & 1.079704e-07 & 1\\ 
 &  & 4.936912e+04 & 4.397571e-01 & -2.021232e+01 & 4.482975e-04 & -2.596585e-08 & \\ 
\ch{Co3O4(s)} & Cobalt Oxide & 3.805484e+05 & -1.311011e+01 & -4.889755e+01 & 1.492909e-02 & -1.851336e-06 & 1\\ 
\ch{CoCl2(s,l)} & Cobalt Chloride & 1.182421e+05 & -3.463875e-01 & -4.766135e+01 & 3.411071e-03 & -9.200809e-07 & 1\\ 
 &  & 1.126187e+05 & -1.194465e+00 & -3.642867e+01 & 2.978683e-03 & -3.093353e-07 & \\ 
\ch{CoF2(s,l)} & Cobalt Fluoride & 1.506989e+05 & -3.193462e+00 & -3.181019e+01 & 6.137417e-03 & -1.223355e-06 & 1\\ 
 &  & 1.485934e+05 & 4.552908e+00 & -8.005494e+01 & -1.340822e-04 & 1.341266e-08 & \\ 
\ch{CoF3(s)} & Cobalt Fluoride & 1.750972e+05 & -1.031290e-01 & -6.748535e+01 & 1.091077e-03 & -6.583164e-08 & 1\\ 
\ch{CoO(s)} & Cobalt Oxide & 1.100576e+05 & 5.701443e-01 & -3.845427e+01 & 1.455438e-04 & 5.439530e-08 & 1\\ 
\ch{CoSO4(s)} & Cobalt Sulfate & 3.090612e+05 & -1.116262e+01 & -3.771972e+01 & 1.342978e-02 & -1.891004e-06 & 1\\ 
\ch{Cr(s,l)} & Chromium & 4.776074e+04 & -4.402638e-01 & -1.579587e+01 & 1.135168e-03 & -5.798430e-08 & 1\\ 
 &  & 4.812046e+04 & 4.364483e+00 & -4.903639e+01 & -8.052095e-04 & 2.749278e-08 & \\ 
\ch{Cr23C6(s)} & Chromium Carbide & 1.651885e+06 & -2.815593e+01 & -3.629898e+02 & 5.382902e-02 & -7.115388e-06 & 1\\ 
\ch{Cr2N(s)} & Chromium Nitride & 1.676501e+05 & -3.745241e-01 & -5.139066e+01 & 2.217768e-03 & -1.117960e-07 & 1\\ 
\ch{Cr2O3(s,l)} & Chromium Oxide & 3.209921e+05 & -4.756124e+00 & -6.173275e+01 & 6.743034e-03 & -8.731748e-07 & 1\\ 
 &  & 3.151896e+05 & 7.173936e+00 & -1.402790e+02 & -5.100608e-04 & -4.577192e-09 & \\ 
\ch{Cr3C2(s)} & Chromium Carbide & 3.249597e+05 & -5.902197e+00 & -5.588923e+01 & 9.151586e-03 & -1.088919e-06 & 1\\ 
\ch{Cr7C3(s)} & Chromium Carbide & 6.109513e+05 & -9.093707e+00 & -1.268842e+02 & 1.555679e-02 & -1.704735e-06 & 1\\ 
\ch{CrN(s)} & Chromium Nitride, Carlsbergite & 1.187703e+05 & -4.973699e-01 & -3.255611e+01 & 1.795595e-03 & -2.412700e-07 & 1\\ 
\ch{Cu(OH)2(s)} & Copper Hydroxide & 2.063558e+05 & -3.959163e+00 & -4.948256e+01 & 4.803361e-03 & -6.293209e-07 & 1\\ 
\ch{Cu(s,l)} & Copper & 4.063664e+04 & -1.796348e-01 & -1.541808e+01 & 1.094894e-03 & -1.863786e-07 & 1\\ 
 &  & 3.992359e+04 & 1.143037e+00 & -2.354187e+01 & 2.254564e-04 & -2.859314e-08 & \\ 
\ch{Cu2O(s,l)} & Copper Oxide & 1.312161e+05 & -2.985239e+00 & -3.107042e+01 & 5.111241e-03 & -8.064332e-07 & 1\\ 
 &  & 9.760053e+04 & -4.553264e+01 & 2.730661e+02 & 2.716321e-02 & -2.449896e-06 & \\ 
\ch{CuCN(s)} & Copper Cyanide & 1.716651e+05 & -3.780381e+00 & -2.525504e+01 & 7.348382e-03 & -1.652732e-06 & 1\\ 
\ch{CuCl(s,l)} & Copper Chloride & 7.166240e+04 & -2.279191e+00 & -1.794801e+01 & 7.232716e-03 & -1.983595e-06 & 1\\ 
 &  & 7.164885e+04 & 2.401760e+00 & -4.473335e+01 & 3.431803e-04 & -3.623642e-08 & \\ 
\ch{CuCl2(s)} & Copper Chloride & 9.455309e+04 & -6.600209e-01 & -4.377040e+01 & 2.643987e-03 & -4.290951e-07 & 1\\ 
\ch{CuF(s)} & Copper Fluoride & 8.410522e+04 & 2.339656e-01 & -3.397521e+01 & 1.885845e-03 & -2.357838e-07 & 1\\ 
\ch{CuF2(s,l)} & Copper Fluoride & 1.239210e+05 & -4.086542e+00 & -2.582427e+01 & 7.893781e-03 & -1.490332e-06 & 1\\ 
 &  & 1.201494e+05 & 2.897313e+00 & -6.539023e+01 & 9.315675e-04 & -8.899353e-08 & \\ 
\ch{CuO(s)} & Copper Oxide & 8.895417e+04 & -2.221543e+00 & -2.134043e+01 & 3.694321e-03 & -5.357891e-07 & 1\\ 
\ch{CuSO4(s)} & Copper Sulfate & 2.840470e+05 & -1.142242e+01 & -3.544105e+01 & 1.361580e-02 & -1.928828e-06 & 1\\ 
\ch{Fe(CO)5(l)} & Iron Carbonyl & 7.203824e+05 & -3.761358e+01 & 1.681905e+01 & 1.282803e-01 & -6.092806e-05 & 1\\ 
\ch{Fe(OH)2(s)} & Iron Hydroxide & 2.304052e+05 & -4.869319e+00 & -4.809856e+01 & 5.794217e-03 & -7.697445e-07 & 1\\ 
\ch{Fe(OH)3(s)} & Iron Hydroxide & 3.161186e+05 & -1.177456e+01 & -3.624228e+01 & 1.140455e-02 & -1.475673e-06 & 1\\ 
  \hline
 \multicolumn{8}{p{14cm}}{\textbf{References.} (1) \citet{Chase1998}; (2) \citet{Moses1992Icar...99..318M}, \citet{Prydz1972}, NIST Chemistry Webbook; (3) \citet{Goodwin1985JPCRD..14..849G}; (4) \citet{yaws1999chemical}, NIST Chemistry Webbook; (5) \citet{Sharp1990ApJS...72..417S}; (6) \citet{Murphy2005}, \citet{Wagner2008}; (7) \citet{LandoltBornstein2001:sm_lbs_978-3-540-45367-3_3}; (8) \citet{Lide2009crc}, \citet{Haar1978JPCRD...7..635H};  (9) \citet{Gail2013AA...555A.119G}; (10) \citet{Barin1995};}
  \end{tabular}
\end{table*}  
  
\begin{table*}
  \contcaption{An overview of condensates and the fit coefficients for the equilibrium constants included in FastChem.}
  \centering
  \scriptsize
  \begin{tabular}{llcccccc}
  \hline\hline
  Species              & Name                      & $a_0$       & $a_1$        & $b_0$        &  $b_1$      &  $b_2$       & References\\
  \hline
\ch{Fe(s,l)} & Iron & 4.974655e+04 & -1.591894e+00 & -9.409153e+00 & 2.776344e-03 & -2.676971e-07 & 1\\ 
 &  & 5.079703e+04 & 3.096658e+00 & -4.091199e+01 & -4.838448e-05 & -7.731235e-09 & \\ 
\ch{Fe2(SO4)3(s)} & Iron Sulfate & 8.653884e+05 & -2.297044e+01 & -1.590785e+02 & 2.277154e-02 & -2.696870e-06 & 1\\ 
\ch{Fe2O3(s)} & Iron Oxide, Hematite & 2.875123e+05 & -8.618740e+00 & -4.016576e+01 & 1.291300e-02 & -1.814306e-06 & 1\\ 
\ch{Fe2SiO4(s)} & Fayalite & 4.534231e+05 & -1.015392e+00 & -1.192593e+02 & 2.752740e-03 & -3.578755e-08 & 5\\ 
\ch{Fe3O4(s)} & Iron Oxide, Magnetite & 4.027661e+05 & -9.466152e+00 & -6.950241e+01 & 1.508513e-02 & -1.944451e-06 & 1\\ 
\ch{FeCl2(s,l)} & Iron Chloride & 1.203448e+05 & -3.734284e-01 & -4.609400e+01 & 2.428654e-03 & -4.403345e-07 & 1\\ 
 &  & 1.133134e+05 & -5.693037e+00 & -6.086630e+00 & 6.792618e-03 & -7.496602e-07 & \\ 
\ch{FeCl3(s,l)} & Iron Chloride & 1.414220e+05 & -3.683131e+00 & -4.480520e+01 & 1.026354e-02 & -3.135510e-06 & 1\\ 
 &  & 1.360565e+05 & -3.706127e+00 & -3.497542e+01 & 8.553835e-03 & -1.302368e-06 & \\ 
\ch{FeF2(s,l)} & Iron Fluoride & 1.533412e+05 & -3.671062e+00 & -2.813476e+01 & 6.031207e-03 & -1.077456e-06 & 1\\ 
 &  & 1.503731e+05 & 2.114308e+00 & -6.271308e+01 & 9.913520e-04 & -8.604572e-08 & \\ 
\ch{FeF3(s)} & Iron Fluoride & 2.039979e+05 & 6.578156e-02 & -6.800792e+01 & 6.230488e-04 & 6.927721e-08 & 1\\ 
\ch{FeO(s,l)} & Iron Oxide & 1.125810e+05 & -8.222241e-01 & -2.941345e+01 & 2.022018e-03 & -2.157028e-07 & 1\\ 
 &  & 1.125140e+05 & 3.159906e+00 & -5.608247e+01 & -8.454633e-06 & -1.060767e-08 & \\ 
\ch{FeS(s,l)} & Troilite & 9.509604e+04 & -4.082122e+00 & -1.321710e+01 & 1.163709e-02 & -2.944081e-06 & 1\\ 
 & Iron Sulfide & 9.213673e+04 & 2.251085e+00 & -4.666790e+01 & 9.905113e-05 & -2.071871e-08 & \\ 
\ch{FeS2(s)} & Iron Sulfide, Pyrite & 1.362424e+05 & -5.216929e+00 & -2.472416e+01 & 7.246209e-03 & -1.224809e-06 & 1\\ 
\ch{FeSO4(s)} & Iron Sulfate & 3.124384e+05 & -1.174652e+01 & -3.362834e+01 & 1.366547e-02 & -1.903293e-06 & 1\\ 
\ch{H2O(s,l)} & Water & 1.161044e+05 & -3.718075e+00 & -9.969769e+00 & -3.487074e-02 & 5.802833e-05 & 6\\ 
 &  & 1.159205e+05 & -1.223011e+01 & 2.278135e+01 & 4.478035e-02 & -2.373704e-05 & \\ 
\ch{H2SO4 * 2 H2O(s,l)} & Sulfuric Acid, Dihydrate & 5.421586e+05 & 0.000000e+00 & -1.859508e+02 & 0.000000e+00 & 0.000000e+00 & 1\\ 
 &  & 5.379818e+05 & -2.361357e+01 & -4.825948e+01 & 3.685278e-02 & -4.514038e-06 & \\ 
\ch{H2SO4 * 3 H2O(s,l)} & Sulfuric Acid, Trihydrate & 6.600920e+05 & 0.000000e+00 & -2.255215e+02 & 0.000000e+00 & 0.000000e+00 & 1\\ 
 &  & 6.553913e+05 & -2.443345e+01 & -8.133285e+01 & 3.702767e-02 & -5.141492e-06 & \\ 
\ch{H2SO4 * 4 H2O(s,l)} & Sulfuric Acid, Tetrahydrate & 7.776671e+05 & 0.000000e+00 & -2.645269e+02 & 0.000000e+00 & 0.000000e+00 & 1\\ 
 &  & 7.737246e+05 & -1.921562e+01 & -1.506327e+02 & 2.900250e-02 & -3.713059e-06 & \\ 
\ch{H2SO4 * H2O(s,l)} & Sulfuric Acid, Monohydrate & 4.241238e+05 & 0.000000e+00 & -1.476872e+02 & 0.000000e+00 & 0.000000e+00 & 1\\ 
 &  & 4.207448e+05 & -1.683880e+01 & -4.909595e+01 & 2.826498e-02 & -4.992257e-06 & \\ 
\ch{H3PO4(s,l)} & Phosphoric Acid & 3.649398e+05 & -3.610999e+02 & 1.635643e+03 & 1.659642e+00 & -1.244896e-03 & 1\\ 
 &  & 3.858749e+05 & -1.943246e+01 & -5.302719e+00 & 2.691054e-02 & -2.100131e-06 & \\ 
\ch{K(HF2)(s,l)} & Potassium Fluoride & 1.686015e+05 & 1.626319e+01 & -1.335623e+02 & -8.347054e-02 & 5.771035e-05 & 1\\ 
 &  & 1.665289e+05 & 8.874422e-01 & -6.179823e+01 & 1.281717e-03 & -1.627841e-07 & \\ 
\ch{K(s,l)} & Potassium & 1.077119e+04 & -1.679865e+00 & -4.750111e+00 & 1.105795e-02 & -7.948908e-06 & 1\\ 
 &  & 1.085335e+04 & 1.658842e+00 & -2.136304e+01 & -7.517943e-04 & 1.673343e-07 & \\ 
\ch{K2CO3(s,l)} & Potassium Carbonate & 3.343443e+05 & -8.558037e+00 & -4.687500e+01 & 1.286535e-02 & -1.718329e-06 & 1\\ 
 &  & 3.369816e+05 & 4.157171e+00 & -1.297355e+02 & 3.324597e-03 & -3.044607e-07 & \\ 
\ch{K2O(s)} & Potassium Oxide & 9.572263e+04 & 1.559253e+00 & -5.837793e+01 & 2.141865e-03 & 4.482650e-08 & 1\\ 
\ch{K2O2(s)} & Potassium Peroxide & 1.411746e+05 & -7.040524e-01 & -6.172071e+01 & 4.554021e-03 & -1.252715e-07 & 1\\ 
\ch{K2S(s,l)} & Potassium Sulfide & 1.018381e+05 & 9.711730e+00 & -1.020847e+02 & -1.591247e-02 & 5.117272e-06 & 1\\ 
 &  & 9.804314e+04 & 2.913293e+00 & -6.343656e+01 & 9.016801e-04 & -8.185043e-08 & \\ 
\ch{K2SO4(s,l)} & Potassium Sulfate & 3.454310e+05 & -1.139934e+01 & -4.772189e+01 & 1.574027e-02 & -1.746880e-06 & 1\\ 
 &  & 3.423413e+05 & 5.296285e-01 & -1.169303e+02 & 2.957119e-03 & -2.371220e-07 & \\ 
\ch{K2SiO3(s,l)} & Potassium Silicate & 3.498376e+05 & -1.098489e+01 & -3.601910e+01 & 1.887993e-02 & -3.516949e-06 & 1\\ 
 &  & 3.465636e+05 & 2.172528e+00 & -1.115910e+02 & 2.227290e-03 & -1.900013e-07 & \\ 
\ch{K3Al2Cl9(s)} & Potassium Aluminum Chloride & 5.873654e+05 & 3.236455e-02 & -2.243358e+02 & 7.124397e-03 & -7.854201e-08 & 1\\ 
\ch{K3AlCl6(s)} & Potassium Hexachloroaluminate & 4.113207e+05 & -7.415459e-01 & -1.509520e+02 & 7.704941e-03 & -6.003396e-07 & 1\\ 
\ch{K3AlF6(s)} & Potassium Hexafluoraluminate & 5.284189e+05 & -5.965318e+00 & -1.253129e+02 & 1.265846e-02 & -1.322514e-06 & 1\\ 
\ch{KAlCl4(s)} & Potassium Tetrachloroaluminate & 2.529190e+05 & -1.016468e+00 & -9.212325e+01 & 7.150289e-03 & -6.981276e-07 & 1\\ 
\ch{KAlSi3O8(s)} & Microcline & 9.255717e+05 & -5.767761e-10 & -2.262691e+02 & 3.706558e-13 & -2.893568e-17 & 5\\ 
\ch{KCN(s,l)} & Potassium Cyanide & 1.686987e+05 & 1.120345e+01 & -1.023096e+02 & -2.744777e-02 & 1.017605e-05 & 1\\ 
 &  & 1.665750e+05 & 1.215299e+00 & -4.869149e+01 & 2.327041e-04 & -2.641994e-08 & \\ 
\ch{KCl(s,l)} & Sylvite & 7.799634e+04 & 2.164702e-01 & -3.146200e+01 & 1.372881e-03 & -1.226827e-07 & 1\\ 
 & Potassium Chloride & 7.532773e+04 & 2.964109e-02 & -2.853322e+01 & 2.400027e-03 & -2.560579e-07 & \\ 
\ch{KClO4(s)} & Potassium Perchlorate & 1.941104e+05 & -1.965987e+01 & 1.221054e+01 & 3.800872e-02 & -7.788120e-06 & 1\\ 
\ch{KF(s,l)} & Potassium Fluoride & 8.861397e+04 & -8.456627e-01 & -2.625877e+01 & 2.808236e-03 & -4.810086e-07 & 1\\ 
 &  & 8.727994e+04 & 2.303486e+00 & -4.537539e+01 & 7.036535e-04 & -6.149724e-08 & \\ 
\ch{KH(s)} & Potassium Hydride & 4.356033e+04 & -2.085467e+00 & -1.512901e+01 & 3.606523e-03 & -4.797331e-07 & 1\\ 
\ch{KO2(s)} & Potassium Superoxide & 1.050071e+05 & -7.728754e-01 & -4.074377e+01 & 4.806930e-03 & -8.106701e-07 & 1\\ 
\ch{KOH(s,l)} & Potassium Hydroxide & 1.176667e+05 & -4.819711e-01 & -3.767670e+01 & -7.788299e-03 & 9.191450e-06 & 1\\ 
 &  & 1.164340e+05 & 3.912731e-01 & -4.363535e+01 & 1.633090e-03 & -2.048580e-07 & \\ 
\ch{Mg(OH)2(s)} & Magnesium Hydroxide & 2.389267e+05 & -1.292321e+01 & -5.297023e-01 & 1.971956e-02 & -4.788722e-06 & 1\\ 
\ch{Mg(s,l)} & Magnesium & 1.770507e+04 & -4.329649e-01 & -1.203290e+01 & 1.820868e-03 & -3.855719e-07 & 1\\ 
 &  & 1.695543e+04 & 2.988436e-01 & -1.560396e+01 & 8.906642e-04 & -9.747666e-08 & \\ 
\ch{Mg2C3(s)} & Magnesium Carbide & 2.843523e+05 & -1.188247e+00 & -7.453823e+01 & 2.686587e-03 & -2.338453e-07 & 1\\ 
\ch{Mg2Si(s,l)} & Magnesium Silicide & 9.891257e+04 & -6.845227e-01 & -4.322943e+01 & 2.881142e-03 & -3.449654e-07 & 1\\ 
 &  & 9.100387e+04 & 3.747421e+00 & -6.622348e+01 & 3.256929e-05 & -7.935425e-09 & \\ 
\ch{Mg2SiO4(s,l)} & Forsterite & 4.685245e+05 & -1.260765e+01 & -4.533909e+01 & 1.474156e-02 & -2.019847e-06 & 1\\ 
 & Magnesium Silicate & 4.672008e+05 & 3.105344e+00 & -1.454310e+02 & 1.348332e-03 & -8.110942e-08 & \\ 
\ch{Mg2TiO4(s,l)} & Magnesium Titanium Oxide & 4.700251e+05 & -1.231678e+01 & -4.689025e+01 & 1.565454e-02 & -2.197166e-06 & 1\\ 
 &  & 4.650479e+05 & 6.163277e+00 & -1.646344e+02 & 1.346740e-03 & -1.018401e-07 & \\ 
\ch{Mg3N2(s)} & Magnesium Nitride & 2.218839e+05 & -2.274009e+00 & -6.695963e+01 & 3.991576e-03 & -5.068377e-07 & 1\\ 
\ch{Mg3P2O8(s,l)} & Magnesium Phosphate & 8.208372e+05 & -1.416969e+01 & -1.400854e+02 & 1.687648e-02 & -2.000682e-06 & 1\\ 
 &  & 8.379570e+05 & 2.081898e+01 & -3.891739e+02 & 1.421557e-03 & -1.122682e-07 & \\ 
\ch{MgAl2O4(s,l)} & Spinel & 4.905698e+05 & -1.277185e+01 & -4.605652e+01 & 1.415049e-02 & -1.693535e-06 & 1\\ 
 & Magnesium Aluminum Oxide & 4.747388e+05 & 3.247152e+00 & -1.436686e+02 & 1.810017e-03 & -9.783613e-08 & \\ 
\ch{MgC2(s)} & Magnesium Carbide & 1.794808e+05 & -7.146464e-01 & -4.564324e+01 & 1.609585e-03 & -1.397661e-07 & 1\\ 
\ch{MgCO3(s)} & Magnesium Carbonate & 3.251431e+05 & -1.244718e+01 & -1.310458e+01 & 1.730791e-02 & -3.510841e-06 & 1\\ 
  \hline
  \end{tabular}
\end{table*}

\begin{table*}
  \contcaption{An overview of condensates and the fit coefficients for the equilibrium constants included in FastChem.}
  \centering
  \scriptsize
  \begin{tabular}{llcccccc}
  \hline\hline
  Species              & Name                      & $a_0$       & $a_1$        & $b_0$        &  $b_1$      &  $b_2$       & References\\
  \hline 
\ch{MgCl2(s,l)} & Magnesium Chloride & 1.238737e+05 & -2.052709e+00 & -3.619730e+01 & 5.877005e-03 & -1.489818e-06 & 1\\ 
 &  & 1.201646e+05 & 1.847462e+00 & -5.572742e+01 & 8.343270e-04 & -7.266538e-08 & \\ 
\ch{MgF2(s,l)} & Magnesium Fluoride & 1.712104e+05 & -4.591530e+00 & -2.243934e+01 & 7.438186e-03 & -1.344173e-06 & 1\\ 
 &  & 1.650826e+05 & -1.155217e+00 & -3.845058e+01 & 2.240213e-03 & -1.637437e-07 & \\ 
\ch{MgH2(s)} & Magnesium Hydride & 7.780425e+04 & -6.636259e+00 & -7.191223e-01 & 6.219478e-03 & -8.158665e-07 & 1\\ 
\ch{MgO(s,l)} & Periclase & 1.194249e+05 & -2.372738e+00 & -1.943966e+01 & 2.746009e-03 & -2.913453e-07 & 1\\ 
 & Magnesium Oxide & 1.092185e+05 & -2.366847e+00 & -1.431364e+01 & 1.452149e-03 & -6.711535e-08 & \\ 
\ch{MgS(s)} & Magnesium Sulfide & 9.243983e+04 & -8.157691e-01 & -2.730010e+01 & 1.496078e-03 & -1.497490e-07 & 1\\ 
\ch{MgSO4(s,l)} & Magnesium Sulfate & 3.200432e+05 & -1.284829e+01 & -2.706610e+01 & 1.567552e-02 & -2.606110e-06 & 1\\ 
 &  & 3.242348e+05 & 3.274547e+00 & -1.304625e+02 & 3.745212e-04 & -2.944070e-08 & \\ 
\ch{MgSiO3(s,l)} & Enstatite & 3.458926e+05 & -1.021310e+01 & -2.617338e+01 & 1.218793e-02 & -1.894794e-06 & 1\\ 
 & Magnesium Silicate & 3.308321e+05 & -1.391574e+01 & 1.328227e+01 & 7.824586e-03 & -5.383545e-07 & \\ 
\ch{MgTi2O5(s,l)} & Magnesium Titanium Oxide & 5.804017e+05 & -1.237362e+01 & -6.590303e+01 & 1.405198e-02 & -1.724862e-06 & 1\\ 
 &  & 5.774343e+05 & 9.048780e+00 & -2.071228e+02 & 8.175831e-04 & -9.154633e-08 & \\ 
\ch{MgTiO3(s,l)} & Geikielite & 3.518015e+05 & -9.495908e+00 & -3.177081e+01 & 1.253078e-02 & -1.924395e-06 & 1\\ 
 & Magnesium Titanium Oxide & 3.494273e+05 & 4.621298e+00 & -1.217642e+02 & 8.711160e-04 & -7.260380e-08 & \\ 
\ch{Mn(s,l)} & Manganese & 3.411061e+04 & -2.422703e-01 & -1.619672e+01 & 1.312815e-03 & 3.983368e-08 & 1\\ 
 &  & 3.372262e+04 & 2.270770e+00 & -3.274546e+01 & 3.623250e-04 & -2.876596e-08 & \\ 
\ch{MnS(s)} & Alabandite & 9.064364e+04 & -1.668626e+00 & -1.792366e+01 & 8.069930e-04 & -5.669297e-08 & 5\\ 
\ch{MnSiO3(s)} & Pyroxmangite & 3.328499e+05 & -3.368640e+00 & -6.225336e+01 & 2.116633e-03 & -1.143565e-07 & 5\\ 
\ch{N2(s,l)} & Nitrogen & 1.144108e+05 & 1.180492e+01 & -6.649718e+01 & -1.623163e-01 & 3.501869e-04 & 7\\ 
 &  & 1.142446e+05 & 6.860148e+00 & -4.814188e+01 & -7.092431e-02 & 9.991725e-05 & \\ 
\ch{N2H4(l)} & Hydrazine & 2.112583e+05 & -5.016141e+00 & -4.552837e+01 & 2.328739e-03 & 1.081899e-06 & 1\\ 
\ch{N2O4(s,l)} & Nitrogen Oxide & 2.371881e+05 & 0.000000e+00 & -9.378900e+01 & 0.000000e+00 & 0.000000e+00 & 1\\ 
 &  & 2.153288e+05 & -1.732139e+02 & 8.442573e+02 & 4.729010e-01 & -2.076125e-04 & \\ 
\ch{NH3(s,l)} & Ammonia & 1.356675e+05 & -2.073772e+02 & 8.458273e+02 & 1.637355e+00 & -2.065313e-03 & 8\\ 
 &  & 1.434961e+05 & 1.579249e+00 & -5.407344e+01 & -5.959483e-03 & -4.175931e-06 & \\ 
\ch{NH4Cl(s)} & Ammonium Chloride & 2.120865e+05 & -9.628320e+00 & -2.321631e+01 & 1.028799e-02 & -1.594689e-06 & 1\\ 
\ch{NH4ClO4(s)} & Ammonium Perchlorate & 3.268847e+05 & -2.461177e+01 & -7.585352e-01 & 2.965828e-02 & -4.378953e-06 & 1\\
\ch{Na(s,l)} & Sodium & 1.293288e+04 & -1.665316e+00 & -5.196019e+00 & 9.504856e-03 & -6.021013e-06 & 1\\ 
 &  & 1.304738e+04 & 1.729799e+00 & -2.260327e+01 & -8.367080e-04 & 1.567141e-07 & \\ 
\ch{Na2O(s,l)} & Sodium Oxide & 1.057691e+05 & -2.150694e+00 & -3.539475e+01 & 4.431860e-03 & -2.177568e-07 & 1\\ 
 &  & 1.018094e+05 & 4.603520e+00 & -7.596361e+01 & 2.082408e-04 & -1.565908e-08 & \\ 
\ch{Na2O2(s)} & Sodium Peroxide & 1.468579e+05 & -4.310953e+00 & -4.014897e+01 & 8.668254e-03 & -1.273752e-06 & 1\\ 
\ch{Na2S(s,l)} & Sodium Sulfide & 1.050216e+05 & 8.653401e+00 & -9.818653e+01 & -1.079498e-02 & 3.035193e-06 & 1\\ 
 &  & 9.800794e+04 & 1.371963e+00 & -5.085516e+01 & 9.718224e-04 & -7.359485e-08 & \\ 
\ch{Na2S2(s,l)} & Sodium Sulfide & 1.384770e+05 & -1.265367e+01 & 1.083078e+01 & 2.874699e-02 & -8.683356e-06 & 1\\ 
 &  & 1.311407e+05 & -1.961069e+01 & 6.892489e+01 & 2.154431e-02 & -3.115168e-06 & \\ 
\ch{Na2SO4(s,l)} & Sodium Sulfate & 3.426922e+05 & -2.253841e+01 & 1.010460e+01 & 4.481540e-02 & -1.098880e-05 & 1\\ 
 &  & 3.434687e+05 & 2.739530e+00 & -1.335028e+02 & 1.716322e-03 & -1.421508e-07 & \\ 
\ch{Na2Si2O5(s,l)} & Sodium Silicate & 5.773671e+05 & -1.983361e+01 & -3.759235e+01 & 2.924649e-02 & -5.317429e-06 & 1\\ 
 &  & 5.762440e+05 & -1.864110e+00 & -1.427231e+02 & 5.951427e-03 & -5.492852e-07 & \\ 
\ch{Na2SiO3(s,l)} & Sodium Silicate & 3.556010e+05 & -1.148912e+01 & -3.428184e+01 & 1.772522e-02 & -3.178852e-06 & 1\\ 
 &  & 3.455189e+05 & -1.006327e+01 & -2.910188e+01 & 8.480798e-03 & -7.291031e-07 & \\ 
\ch{Na3AlCl6(s)} & Sodium Hexachloroaluminate & 4.042589e+05 & -8.119085e-01 & -1.511920e+02 & 6.964421e-03 & -5.080981e-07 & 1\\ 
\ch{Na3AlF6(s,l)} & Cryolite & 5.324731e+05 & -1.328508e+01 & -8.473338e+01 & 2.258069e-02 & -3.148918e-06 & 1\\ 
 &  & 5.394329e+05 & 1.923886e+01 & -3.009206e+02 & 1.504385e-03 & -1.149278e-07 & \\ 
\ch{Na5Al3F14(s,l)} & Chiolite & 1.225776e+06 & -1.886442e+01 & -2.513728e+02 & 3.045841e-02 & -4.387310e-06 & 1\\ 
 &  & 1.254661e+06 & 5.659140e+01 & -7.781243e+02 & 2.631956e-03 & -2.127919e-07 & \\ 
\ch{NaAlCl4(s)} & Sodium Tetrachloroaluminate & 2.484835e+05 & -1.614505e+00 & -8.890879e+01 & 8.297430e-03 & -1.103368e-06 & 1\\ 
\ch{NaAlO2(s)} & Sodium Aluminum Oxide & 2.476933e+05 & -4.963382e+00 & -3.828779e+01 & 5.862325e-03 & -6.164413e-07 & 1\\ 
\ch{NaAlSi3O8(s)} & Albite & 9.228090e+05 & -9.591781e-10 & -2.264186e+02 & 6.208685e-13 & -4.831038e-17 & 5\\ 
\ch{NaCN(s,l)} & Sodium Cyanide & 1.678194e+05 & 3.156953e+00 & -6.260474e+01 & -6.145218e-04 & -9.665279e-07 & 1\\ 
 &  & 1.665917e+05 & 1.130587e+00 & -4.909532e+01 & 5.374873e-04 & -5.037493e-08 & \\ 
\ch{NaCO3(s,l)} & Sodium Carbonate & 3.360592e+05 & -1.103823e+01 & -3.331982e+01 & 1.774692e-02 & -3.027202e-06 & 1\\ 
 &  & 3.371626e+05 & 3.276748e+00 & -1.212668e+02 & 2.500792e-03 & -2.280521e-07 & \\ 
\ch{NaCl(s,l)} & Halite & 7.705306e+04 & -1.494261e-01 & -2.974098e+01 & 1.907680e-03 & -2.831117e-07 & 1\\ 
 & Sodium Chloride & 7.484593e+04 & 2.063974e+00 & -4.190249e+01 & 5.113600e-04 & -4.704151e-08 & \\ 
\ch{NaClO4(s)} & Sodium Perchlorate & 1.926202e+05 & -1.053871e+01 & -4.311559e+01 & 2.536888e-02 & -5.318760e-06 & 1\\ 
\ch{NaF(s,l)} & Sodium Fluoride & 9.145669e+04 & -1.655268e+00 & -2.236020e+01 & 3.733511e-03 & -6.749430e-07 & 1\\ 
 &  & 9.037933e+04 & 3.684347e+00 & -5.574581e+01 & -2.169080e-04 & 6.522889e-09 & \\ 
\ch{NaH(s)} & Sodium Hydride & 4.530074e+04 & -3.496345e+00 & -7.025608e+00 & 5.415961e-03 & -8.554953e-07 & 1\\ 
\ch{NaO2(s)} & Sodium Superoxide & 1.052766e+05 & 5.071979e+00 & -7.457390e+01 & -5.291845e-03 & 1.504676e-06 & 1\\ 
\ch{NaOH(s,l)} & Sodium Hydroxide & 1.196678e+05 & -3.131047e+00 & -2.450604e+01 & 1.339562e-03 & 2.749203e-06 & 1\\ 
 &  & 1.201804e+05 & 3.258577e+00 & -6.415605e+01 & -3.757245e-04 & 5.988784e-09 & \\ 
\ch{Ni(CO)4(l)} & Nickel Carbonyl & 5.934059e+05 & 1.528607e+01 & -2.138743e+02 & -5.414187e-02 & 2.804246e-05 & 1\\ 
\ch{Ni(s,l)} & Nickel & 5.165049e+04 & -7.770138e-01 & -1.418381e+01 & 1.996902e-03 & -3.498124e-07 & 1\\ 
 &  & 5.083042e+04 & 1.209952e+00 & -2.647397e+01 & 2.328150e-04 & -1.132653e-08 & \\ 
\ch{Ni3S2(s,l)} & Heazlewoodite & 2.481926e+05 & 8.649363e+00 & -1.314489e+02 & -3.562614e-02 & 1.910757e-05 & 1\\ 
 & Nickel Sulfide & 2.438967e+05 & 6.973000e+00 & -1.330746e+02 & 1.053581e-03 & -6.722200e-08 & \\ 
\ch{Ni3S4(s)} & Nickel Sulfide & 3.238859e+05 & -5.188410e+00 & -9.446559e+01 & 8.614201e-03 & 1.025834e-07 & 1\\ 
\ch{NiCl2(s,l)} & Nickel Chloride & 1.174571e+05 & -1.285269e+00 & -4.301500e+01 & 3.316884e-03 & -6.381411e-07 & 1\\ 
 &  & 7.809482e+04 & -6.197958e+01 & 3.779006e+02 & 4.270107e-02 & -4.583660e-06 & \\ 
\ch{NiS(s,l)} & Millerite & 9.505018e+04 & -4.413550e+00 & -1.123546e+01 & 8.986445e-03 & -1.742369e-06 & 1\\ 
 & Nickel Sulfide & 9.369626e+04 & 2.420643e+00 & -5.110956e+01 & 6.279077e-04 & -4.356159e-08 & \\ 
\ch{NiS2(s,l)} & Vaesite & 1.339698e+05 & -1.185366e+00 & -4.681556e+01 & 1.960389e-03 & -1.164351e-07 & 1\\ 
 & Nickel Sulfide & 1.028536e+05 & -4.665695e+01 & 2.700304e+02 & 3.175018e-02 & -3.373018e-06 & \\ 
\ch{O2S(OH)2(s,l)} & Sulfuric Acid & 3.035777e+05 & 0.000000e+00 & -1.060685e+02 & 0.000000e+00 & 0.000000e+00 & 1\\ 
 &  & 3.006315e+05 & -1.618224e+01 & -1.157076e+01 & 2.471556e-02 & -3.358190e-06 & \\ 
  \hline
  \end{tabular}
\end{table*}

\begin{table*}
  \contcaption{An overview of condensates and the fit coefficients for the equilibrium constants included in FastChem.}
  \centering
  \scriptsize
  \begin{tabular}{llcccccc}
  \hline\hline
  Species              & Name                      & $a_0$       & $a_1$        & $b_0$        &  $b_1$      &  $b_2$       & References\\
  \hline  
\ch{P(s,l)} & Phosphorus & 3.877067e+04 & 8.852251e+00 & -5.893706e+01 & -3.585460e-02 & 2.356430e-05 & 1\\ 
 &  & 3.813711e+04 & 5.246809e-01 & -1.802322e+01 & 1.462641e-04 & -2.276066e-08 & \\
\ch{(P2O5)2(s)} & Phosphorus Oxide & 8.071620e+05 & -3.453923e+01 & -3.775233e+01 & 4.366846e-02 & -5.766348e-06 & 1\\ 
\ch{P3N5(s)} & Phosphorus Nitride & 4.356591e+05 & -1.015124e+01 & -7.275974e+01 & 2.070140e-02 & -3.157142e-06 & 1\\ 
\ch{P4S3(s,l)} & Phosphorus Sulfide & 2.792320e+05 & 0.000000e+00 & -1.151194e+02 & 0.000000e+00 & 0.000000e+00 & 1\\ 
 &  & 2.796414e+05 & 2.959176e+00 & -1.345227e+02 & 1.085772e-03 & -1.263153e-07 & \\ 
\ch{S(s,l)} & Sulfur & 3.324780e+04 & 8.230502e-01 & -1.915434e+01 & -7.473219e-03 & 6.541973e-06 & 1\\ 
 &  & 3.414358e+04 & 3.645266e+00 & -3.927119e+01 & -2.558100e-03 & 4.344450e-07 & \\ 
\ch{SCl2(l)} & Sulfur Chloride & 6.936735e+04 & 3.233471e+00 & -5.894255e+01 & -1.073982e-03 & 3.844648e-07 & 1\\
\ch{Si(s,l)} & Silicon & 5.377075e+04 & -1.906343e+00 & -6.697187e+00 & 2.855366e-03 & -5.042181e-07 & 1\\ 
 &  & 4.851793e+04 & 8.361541e-01 & -2.045829e+01 & -6.642873e-05 & 1.825015e-09 & \\ 
\ch{Si3N4(s)} & Silicon Nitride, Alpha & 4.772060e+05 & -9.361923e+00 & -6.235464e+01 & 7.210523e-03 & -6.175834e-07 & 1\\ 
\ch{SiC(s)} & Silicon Carbide, Beta & 1.482157e+05 & -3.559140e+00 & -1.477158e+01 & 2.955111e-03 & -2.645741e-07 & 1\\ 
\ch{SiO(s)} & Silicon Oxide & 1.451320e+05 & -1.910140e+00 & -2.029867e+01 & 6.185105e-04 & -4.216349e-08 & 9\\ 
\ch{SiO2(s,l)} & Silicon Dioxide & 2.225032e+05 & -5.492533e+00 & -2.075544e+01 & 6.331284e-03 & -9.105538e-07 & 10\\ 
 &  & 2.185727e+05 & -6.967639e+00 & -5.420608e+00 & 3.986627e-03 & -2.747098e-07 & \\ 
\ch{SiS2(s,l)} & Silicon Sulfide & 1.465184e+05 & -2.740423e-01 & -5.035944e+01 & 2.283212e-03 & -2.893822e-07 & 1\\ 
 &  & 1.231541e+05 & -3.994238e+01 & 2.255866e+02 & 2.545788e-02 & -2.489610e-06 & \\ 
\ch{Ti(s,l)} & Titanium & 5.677155e+04 & -1.227259e+00 & -1.096439e+01 & 2.166856e-03 & -2.322204e-07 & 1\\ 
 &  & 5.977556e+04 & 4.249729e+00 & -4.990116e+01 & -3.932996e-04 & 2.304430e-09 & \\ 
\ch{Ti2O3(s,l)} & Titanium Oxide & 3.847074e+05 & -1.095307e+01 & -2.764763e+01 & 1.664392e-02 & -2.602469e-06 & 1\\ 
 &  & 3.750538e+05 & 2.502303e+00 & -1.047853e+02 & 1.357783e-03 & -1.225017e-07 & \\ 
\ch{Ti3O5(s,l)} & Titanium Oxide & 6.136540e+05 & -1.483965e+01 & -5.925200e+01 & 2.465791e-02 & -4.287191e-06 & 1\\ 
 &  & 6.049721e+05 & 8.774242e+00 & -2.040415e+02 & 1.095428e-03 & -1.227866e-07 & \\ 
\ch{Ti4O7(s,l)} & Titanium Oxide & 8.457487e+05 & -8.810282e+00 & -1.475694e+02 & 1.359895e-02 & -1.740277e-06 & 1\\ 
 &  & 8.376337e+05 & 1.408512e+01 & -2.981153e+02 & 8.591056e-04 & -1.320712e-07 & \\ 
\ch{TiC(s,l)} & Titanium Carbide & 1.645199e+05 & -3.261366e+00 & -1.755256e+01 & 3.346973e-03 & -3.490044e-07 & 1\\ 
 &  & 1.633484e+05 & 6.162907e+00 & -8.301350e+01 & -1.081446e-03 & 3.146153e-08 & \\ 
\ch{TiCl2(s)} & Titanium Chloride & 1.478096e+05 & -2.010910e+00 & -3.934015e+01 & 3.484307e-03 & -3.868436e-07 & 1\\ 
\ch{TiCl3(s)} & Titanium Chloride & 1.878227e+05 & 7.714631e-01 & -6.992851e+01 & 3.094332e-04 & 1.335437e-07 & 1\\ 
\ch{TiCl4(s,l)} & Titanium Chloride & 2.125171e+05 & 0.000000e+00 & -7.629495e+01 & 0.000000e+00 & 0.000000e+00 & 1\\ 
 &  & 2.131810e+05 & 3.813615e+00 & -1.000270e+02 & -3.217674e-05 & 2.045416e-07 & \\ 
\ch{TiF3(s)} & Titanium Fluoride & 2.580094e+05 & -1.644778e+00 & -5.896958e+01 & 2.724400e-03 & -1.479247e-07 & 1\\ 
\ch{TiF4(s)} & Titanium Fluoride & 2.923188e+05 & -7.799828e+00 & -3.748423e+01 & 1.435265e-02 & -2.549007e-06 & 1\\ 
\ch{TiH2(s)} & Titanium Hydride & 1.249501e+05 & -8.105000e+00 & 4.428271e+00 & 7.428109e-03 & -9.777304e-07 & 1\\ 
\ch{TiN(s,l)} & Titanium Nitride & 1.537865e+05 & -2.703635e+00 & -1.981151e+01 & 3.014155e-03 & -3.107625e-07 & 1\\ 
 &  & 1.307269e+05 & -1.079465e+01 & 5.022860e+01 & 3.293668e-03 & -1.553395e-07 & \\ 
\ch{TiO(s,l)} & Titanium Oxide & 1.514670e+05 & -3.675177e+00 & -1.482323e+01 & 4.743702e-03 & -5.181798e-07 & 1\\ 
 &  & 1.491392e+05 & 3.610219e+00 & -6.116719e+01 & -2.254299e-04 & -7.164901e-09 & \\ 
\ch{TiO2(s,l)} & Rutile & 2.293679e+05 & -5.579366e+00 & -2.061548e+01 & 6.888392e-03 & -1.023248e-06 & 1\\ 
 & Titanium Oxide & 2.279416e+05 & 2.995753e+00 & -7.643098e+01 & 4.869790e-04 & -4.665479e-08 & \\ 
\ch{V(s,l)} & Vanadium & 6.178073e+04 & -1.058384e+00 & -1.210524e+01 & 1.460874e-03 & -1.709780e-07 & 1\\ 
 &  & 6.258078e+04 & 1.907257e+00 & -3.350554e+01 & 3.464223e-04 & -2.916535e-08 & \\ 
\ch{V2O3(s,l)} & Karelianite & 3.594059e+05 & -4.525272e+00 & -6.248195e+01 & 5.547256e-03 & -5.116554e-07 & 1\\ 
 & Vanadium Oxide & 3.430939e+05 & -2.182166e+00 & -6.895187e+01 & 2.726860e-03 & -1.688924e-07 & \\ 
\ch{V2O4(s,l)} & Vanadium Oxide & 4.138200e+05 & -7.706466e+00 & -6.324071e+01 & 1.428494e-02 & -2.749633e-06 & 1\\ 
 &  & 4.006761e+05 & -7.197933e+00 & -5.418588e+01 & 7.151172e-03 & -5.011278e-07 & \\ 
\ch{V2O5(s,l)} & Shcherbinaite & 4.580524e+05 & -1.240157e+01 & -5.194057e+01 & 1.863388e-02 & -3.815827e-06 & 1\\ 
 & Vanadium Oxide & 4.536238e+05 & 4.659504e-01 & -1.230467e+02 & 2.126091e-03 & -1.807118e-07 & \\ 
\ch{VN(s)} & Vanadium Nitride & 1.443584e+05 & -2.457896e+00 & -2.053464e+01 & 2.367187e-03 & -1.939530e-07 & 1\\ 
\ch{VO(s,l)} & Vanadium Oxide & 1.432838e+05 & -3.129339e+00 & -1.797255e+01 & 4.472298e-03 & -6.021079e-07 & 1\\ 
 &  & 1.345012e+05 & -1.960904e+00 & -1.874037e+01 & 1.567102e-03 & -1.023190e-07 & \\ 
\ch{Zn(s,l)} & Zinc & 1.575906e+04 & -1.232552e-01 & -1.428515e+01 & 1.436668e-03 & -3.507159e-07 & 1\\ 
 &  & 1.527125e+04 & 1.050780e+00 & -2.053114e+01 & 1.544112e-04 & -1.683046e-08 & \\ 
\ch{ZnSO4(s)} & Zinc Sulfate & 2.841321e+05 & -1.432225e+01 & -1.911778e+01 & 2.033870e-02 & -3.227622e-06 & 1\\ 
  \hline
  \end{tabular}
\end{table*}

\section{Element-condensate mapping}
\label{sec:puzzle}

In order to determine the correlation between the condensed species with their associated elements even at the highest number of possible stable condensates according to the phase rule one can apply a simple scheme that is briefly discussed in the following. The scheme is applied to the outermost region of the protoplanetary disk from section~\ref{ssec:proto_disk} and its results are summarised in Table~\ref{tab:condensates_disk}.

\begin{figure*}
    \centering
	\resizebox{0.8\textwidth}{!}{
    \includegraphics{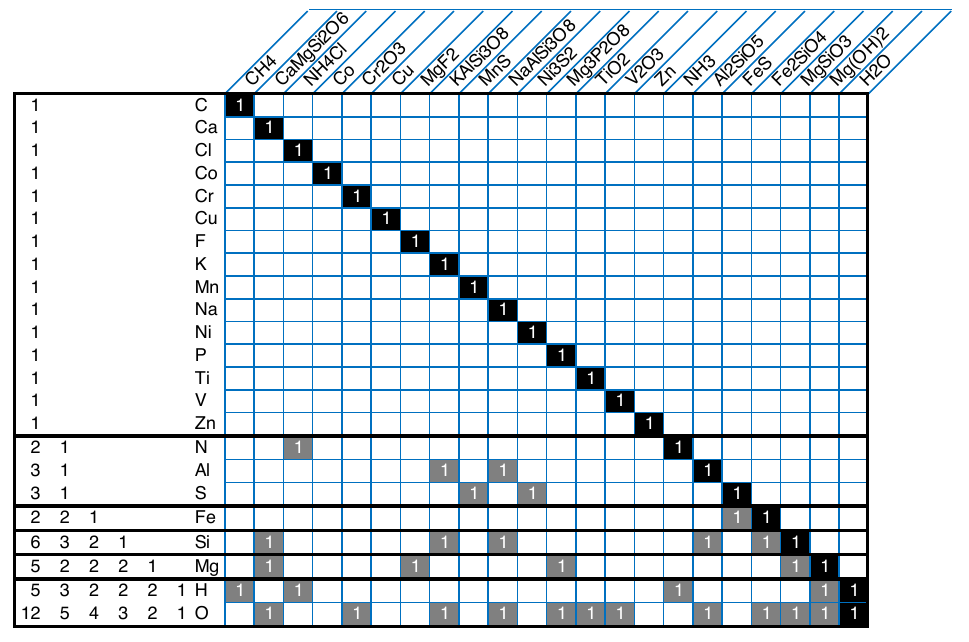}}
	\caption{Mapping between the chemical elements and their corresponding condensates according to the phase rule (see text for details).}
	\label{fig:disk_puzzle}
\end{figure*}

Figure~\ref{fig:disk_puzzle} depicts the end-result of how a unique mapping between the chemical elements and their corresponding condensates according to the phase rule can be found. The right hand side of the figure defines a matrix, which entries are ``1'', if the condensate contains the respective element. The number in the outermost left column on the left hand side of the figure indicates how many condensates an element can be associated with. The elements are then sorted by this number, e.g. carbon receives a ``1'', because the only condensate containing carbon is \ch{CH4}, nitrogen receives a ``2'', because \ch{NH4Cl} and \ch{NH3} include nitrogen, and so on. The mapping of the elements \ch{C},~\ch{Ca},~\ldots,~\ch{Zn} is already determined, since they have only one condensate to associate with. These condensates are therefore not available to other elements. In the next step, the numbers on the left hand side in Fig.~\ref{fig:disk_puzzle} are updated accordingly (second column). The procedure is repeated until no condensate is left. 

In our example the final condensate is \ch{H2O} which can be assigned to either hydrogen or oxygen, since the total number of elements exceeds the total number of condensates by one. If a bijective mapping is desired, a secondary criterion must be utilised, such as for example the stoichiometrically normalised element abundance. In this case, we would assign oxygen, the less abundant element, to \ch{H2O}. Note, that the domain of the mapping must then be reduced by hydrogen which is left over. The described algorithm works because the condensates are linear independent (see sections~\ref{sec:gibbs_phase_rule} and \ref{sec:fastchem_cond}).

\bsp	
\label{lastpage}
\end{document}